# ProdPilot: A Layered Model Context Protocol System for Automated Production Readiness and Deployment of Vibe-Coded Projects

Muhammad Sudais Khalid[1,2#], Muhammad Junaid Asif [1,3*], Rehan Naveed Abbasi[2], and Rana Fayyaz Ahmad [1]
[1] Natural Language Processing Lab, Artificial Intelligence Technology Centre (AITeC), Islamabad (44000), Pakistan
[2] Department of Artificial Intelligence, Shifa Tameer-e-Millat University, Islamabad (44000), Pakistan
[3] Faculty of IT and Computer Sciences, University of Central Punjab (UCP) Lahore (54000), Pakistan
e-mails: # bsai-23f-0050@stmu.edu.pk, * junaid.asif@ncp.edu.pk

***Abstract*—AI-assisted development environments turn natural-language requests into working applications, yet the projects they produce often fail on their first deployment for operational reasons: credentials in source files, missing security middleware, no container definition, no health endpoint and no delivery pipeline. Automated repair of such defects usually relies on free-form model generation, which varies between runs and is seldom checked independently. This paper presents ProdPilot, a local Model Context Protocol server that audits Node.js and React projects against 50 production readiness rules, repairs violations through a bounded loop, gates the result and deploys it. Each rule is classified by what its fix requires: 28 fixes are fixed templates, 16 are filled with values extracted from the project and only 6 are written by the IDE agent's model, so 88.0% of the ruleset needs no model-written content, and across 499 real repositories 93.5% of the violations found fall in those classes. A fix, however it was produced, is accepted only if the checker that flagged the rule now passes. In a full-chain run over 23 sample projects, template and extraction fixes that were applied as written passed their checkers in 87.6% and 95.0% of cases, and independent verification rejected 25 of the 237 fixes reported as applied. A deployability model trained on 684 real deployments ranks held-out projects at 0.909 ROC AUC, a gated project was deployed, smoke-tested and connected to an active CI/CD pipeline in 82 seconds, and the unchanged server worked from the IDE agents in VS Code, Cursor and Windsurf, where live agents resolved all six delegated rules on the first attempt under the same verification. ProdPilot itself runs no model.**

## I. Introduction

AI-assisted development environments such as VS Code with GitHub Copilot, Cursor and Windsurf now turn natural-language requests into running applications. This way of building software, termed vibe coding, moves the decision of how a system is implemented from the developer to the model and leaves open who is accountable for the result [1]. A systematic review of 44 primary studies finds that the productivity gains of AI code assistants come together with measurable security risks [2].

The weaknesses of generated code tend to appear late. Backend code from current models has been found weaker in quality and in architecture even when it works [3], and an analysis of three million application reviews shows model errors reaching end users instead of being caught during development [4]. At deployment the dominant failures are operational rather than logical: credentials in source files, missing security middleware, no container definition, no health endpoint and no delivery pipeline. None of these matters on the developer's machine, so they surface together on the first real server. The developers producing such projects are also unlikely to have operations training; only about five percent of organisations provide formal DevOps training [5].

Existing work treats automated remediation as a generation problem. Program repair adapts code models to produce patches for classes of defects [6], [7], and a second line of work wraps the model in a feedback loop of tests or static analysis that returns faulty output for correction [8]–[11]. Because generated output varies from run to run, work on hallucination argues for deterministic limits around the model [12]–[14]. What this literature does not measure is how much of a remediation task needs a model at all.

Production readiness is a domain where that question has a useful answer. Many requirements are the same for every project, so their fixes can be fixed templates. Others differ only in values already present in the codebase, such as the environment variable that should replace a hardcoded password, and can be filled in by extraction. Only fixes that depend on how a particular codebase is organised need to be written by a model. Those fixes need not be trusted either: each is accepted only when the check that flagged the problem passes again.

No existing tool covers the path from detection to a verified deployment. Scanners report findings without repairing them, and their output is inconsistent enough to need reconciliation before it can drive automation [15], [16]. Pipeline frameworks generate and tune delivery artifacts but neither audit the project nor show that the pipeline they produce is live [17]–[20]. Readiness frameworks remain qualitative [21], DevSecOps surveys catalogue practices without linking findings to deployment failure [22], [23], and published Model Context Protocol (MCP) systems use the protocol to give a model they control access to tools [24], [25].

This paper presents ProdPilot, a local MCP server that covers that path for Node.js projects with Express and React projects with Vite. It holds no model: the IDE's own agent applies every edit, and ProdPilot supplies the rules, the fix contracts, the verification, a deployability estimate and the deployment pipeline. The work is guided by five research questions:

**RQ1:** To what extent can a rule-classification approach across static, parametric and delegated fix classes minimise reliance on free-form model generation while maintaining high fix accuracy in automated production readiness remediation?

**RQ2:** What proportion of real-world production readiness violations in AI-generated Node.js and React code-bases can be resolved through deterministic means, either fixed templates or mechanical extraction, without any model-authored fix?

**RQ3:** Does independent re-verification of every fix, regardless of how it was produced, catch a meaningfully higher rate of incorrect or incomplete fixes than trusting agent self-report?

**RQ4:** Can deployment success be predicted more accurately from static audit signals using a model trained on real deployment outcomes than by treating rule compliance itself as the ground truth?

**RQ5:** Does a protocol-based integration layer achieve reliable production readiness automation across multiple AI-powered IDEs without requiring IDE-specific implementation?

This paper contributes the following:

- A classification of production readiness fixes by what producing them requires. Of the 50 implemented rules, 44 (88.0%) are repaired without model-written content, a share computed by the rule store itself, and 93.5% of the violations found across 499 real repositories fall in those classes.
- Verification that ignores authorship. A fix from a template, an extraction or the agent is accepted only if the flagging checker passes on the files on disk, and a fix that breaks a previously passing rule is reverted.
- A reversal of the usual MCP arrangement: the tool server is the deterministic side, and the only model in the system belongs to the client.
- A deployability model trained on 684 real deployments instead of rule compliance, constrained so that a repair can only keep or raise its estimate, and used beside the audit in a release gate.
- A deployment pipeline that creates the service before wiring CI/CD and was run end to end on real infrastructure.

ProdPilot is fully implemented, and the results reported here come from real deployments rather than simulation.

The remaining sections are arranged as follows. Section II introduces the concepts the work builds on. Section III reviews related work, paper by paper, with the gap each leaves and how ProdPilot addresses it. Section IV states the problem and the design challenges. Section V describes the architecture of ProdPilot layer by layer, and ection VI explains how it was built and tested. Section VII reports the evaluation, and Section VIII concludes.

## II. Theoretical Background

This section sets out the concepts that the rest of the paper relies on: how AI-assisted development produces code, what a deployment pipeline expects of that code, how static analysis and program repair work, why generated output needs to be constrained, how the Model Context Protocol connects tools to an IDE agent, and how deployability can be estimated from data.

### A. Vibe Coding and AI-Assisted Development

Code assistants built on large language models turn a description in natural language into source code, and in agent mode they also edit files, run commands and call external tools for the developer. Meske et al. describe the resulting practice as a change in how intent is mediated: the developer states goals in dialogue, the model infers an implementation, and expertise moves from writing code towards directing and checking it [1]. Two properties of this setting shape the rest of this paper. The model is probabilistic, so the same request can yield different code on different runs, and the developer often accepts code without reading it closely, so weaknesses in security, configuration and design pass into the project unnoticed [2], [3]. When those weaknesses do surface, they tend to reach users before they reach a reviewer [4].

### B. DevOps, Continuous Delivery and Containers

DevOps joins development and operations through shared practice and automation, and CI/CD, short for continuous integration and delivery, turns each commit into a build, a test run and, when both succeed, a deployment [5], [17]. Containers package an application with its runtime and dependencies so that it behaves the same on a laptop and on a server. A container image is described by a Dockerfile, and Docker is the most widely used tool for building and running such images [17], [26]. Cloud platforms build a project from its Git repository, start the service and check that it answers, usually on a health endpoint [18], [19]. Delivery pipelines for machine learning apply the same ideas to trained models, whose training and serving must also be automated and monitored [20]. For a vibe-coded project the relevant point is that each step depends on artifacts, such as a Dockerfile, an environment template, a health route or a workflow file, that the application does not need in order to run on the developer's machine.

### C. Production Readiness and DevSecOps

Production readiness is the set of properties a service needs, beyond functional correctness, before it can run on shared infrastructure. They include secrets kept out of source and supplied by the environment, security headers and request limits, a reproducible container build, configuration read from the environment, database connections that are pooled, a versioned API with consistent errors, observability through logs, metrics and health checks, and a repository that keeps credentials out of its history. DevSecOps makes the security part of this a continuous concern by adding code, secret and dependency scanning to the delivery pipeline [22], [23]. Scanners are the usual way to check such properties, but their findings differ between tools and include false positives [15]. ProdPilot organises these properties into nine domains and six priority tiers, from P0, where a single failure makes a deployment unsafe, to P5 (Section V).

### D. Static Analysis and Syntax Trees

Static analysis examines code without running it. Some checks need only a file's presence or the lines of a configuration file; others need the structure of the program. An abstract syntax tree represents source code as nested nodes, for example a call to `app.use` with its arguments, and it keeps statements in their original order, so it can answer whether one call comes before another. Signals from several analyses can be combined into one interpretable view of a codebase [16], and static findings can also be returned to a model as guidance towards a corrected program [8], [9]. Secret scanners add a statistical signal: a generated credential is a long string whose characters are close to uniformly distributed, so its Shannon entropy is higher than that of ordinary identifiers.

### E. Program Repair and Model Feedback Loops

Automated program repair takes a faulty program and produces a patch. Recent approaches fine-tune or prompt code models to generate candidate patches, which are then validated against tests [6], [7]. A related design places a general-purpose model inside a loop: the model writes code, a checker such as a test suite, a static analyser or a simulator reports what is wrong, and the report is returned to the model until the check passes or a budget runs out [8], [10], [11]. The checker acts as an oracle. The correctness of such a loop therefore depends on what the oracle can observe, and its cost grows with the number of iterations.

### F. Hallucination and Deterministic Constraints

A hallucination is output that reads fluently but is wrong, invented or inconsistent with its source. Surveys group mitigation techniques into changes at training time, better prompts and inputs, retrieval, quality control after generation, and orchestration across several agents [13], [14]. One family of approaches wraps a model in explicit rules so that its responses stay inside a validated boundary [12]. A further option, the one taken here, is to remove from the model those parts of a task that do not need it, and to check the remaining part against a deterministic test.

### G. The Model Context Protocol

The Model Context Protocol (MCP) standardises how an AI application exchanges context and actions with external programs. A client, such as the agent inside an IDE, connects to a server, discovers the tools it offers and calls them with structured arguments; each result returns as text and as structured content. Messages use JSON-RPC and, for a server on the same machine, travel over standard input and output. Existing systems use MCP to give a model they control access to domain tools, such as vibration analysis for machine maintenance [24] or retrieval over educational sources [25]. Because the protocol is the same for every compliant client, a server written once can in principle serve any IDE that supports it. Each client, however, decides for itself how it starts a server, when it asks the user before running a tool, and how much of a result it passes to its model, which is why cross-client behaviour has to be observed rather than assumed.

### H. Learned Estimation of Deployability

Readiness can be judged by people against staged criteria, as in the technology readiness levels proposed for machine learning systems, where a system moves up a level only after a review [21]. Alternatively, it can be learned from outcomes. Gradient boosting builds an ensemble of shallow decision trees, each fitted to the errors left by the previous ones, and performs well on small tables of numeric features. The histogram-based variant groups feature values into bins, which speeds up training and allows monotonic constraints that force the prediction to move in one direction as a feature grows. The raw scores of a classifier become usable probabilities after calibration, for example by fitting a sigmoid on scores the model has not been trained on. Ranking quality is measured by the area under the ROC curve, which plots the true positive rate against the false positive rate (ROC AUC) and, when positive cases are rare, by the area under the precision-recall curve (PR AUC), whose chance level equals the share of positives rather than one half. Calibration is judged by comparing predicted and observed rates within bins of the estimate, and permutation importance measures how far a metric falls when the values of one feature are shuffled.

## III. Related Work

This section reviews the literature in the thematic groups used throughout the paper. For each study it describes the method and main findings, states the limitation or gap the study leaves, and explains how ProdPilot addresses that gap. Table II and Table III at the end of the section summarise the methodology and gap of every reviewed work.

### A. Generated Code Quality

Meske et al. trace how developers have expressed intent to computers across several decades and, from that history, define vibe coding as a mode of development in which a person and a generative model build software together through dialogue [1]. They identify gains such as wider access to programming and faster delivery, and risks such as codebases nobody fully understands and unclear responsibility for defects, and they close with a research agenda. The contribution is conceptual: the risks are named but no mechanism to contain them is proposed or measured. ProdPilot takes one of those risks, the unclear responsibility for whether generated code is fit to deploy, and makes it concrete as 50 checkable rules, each with a verdict, a location and a named fix.

Zardari et al. screen 40,412 records from five databases and synthesise 44 primary studies published between 2021 and 2025 [2]. They report clear productivity gains in routine coding, documentation and bug fixing, together with recurring security flaws in generated code, among them missing input validation, hardcoded credentials and weak cryptography, and they organise the risks into code-level, reliability and legal or ethical domains. The authors note that human baselines vary between the reviewed studies, and their recommendations rest on careful human review. ProdPilot automates the review for the operational part of the problem: hardcoded credentials, for example, are detected by an entropy scan and repaired by moving the value into the environment, with the result confirmed by the same scan.

Horvat et al. ask two current models to build the same Java and Spring Boot backend through a chain of prompts and measure corrective prompts, manual edits, satisfied requirements, test success and test coverage [3]. One model met 90.25% of the requirements and the other 68.08%, and the two differed markedly in how they layered the application. The comparison covers one application and two models, and it ends at the code and its tests; whether the result can be built, configured and served in production is outside its scope. ProdPilot starts where that evaluation stops: it audits the operational layer, repairs it and proves the outcome with a real build and a live deployment.

Massenon et al. filter three million reviews of 90 AI-powered mobile applications, annotate 1,000 candidates by hand and derive a taxonomy of seven kinds of hallucination that users report [4]. The study is observational and limited to English reviews from two app stores over a fixed period, and the errors it describes are found by users after release. ProdPilot moves detection to before deployment and assigns it to deterministic checkers instead of to the people who use the software.

### B. Protocol-Based Integration

Di Maggio presents an open-source MCP server for machine condition monitoring in which a language model selects and configures deterministic signal-processing tools, such as spectral and envelope analysis and severity assessment against an ISO standard, and then writes a conclusion based on their structured results [24]. The author states that the demonstration data do not represent the variety of real plants, that the workflows are not validated in the field, and that the model's output must stay advisory because it can influence operational decisions. The model is therefore still trusted to choose tools and to synthesise the final answer. ProdPilot never relies on a model's conclusion: every change is judged by the rule's own checker, and most changes require no model at all.

Ho and Lam survey retrieval-augmented generation and MCP in education and propose EduMSRA, an architecture that combines a hierarchical retrieval pipeline, an MCP layer for orchestrating curriculum tools, a module that detects conflicts between sources and a learner model [25]. Their proof of concept runs without language-model generation, without live MCP servers and without human evaluation, so the protocol layer is described rather than exercised with real clients. ProdPilot's protocol layer was exercised end to end from the agents of three IDEs, and the behaviour of each client is recorded in Section VII.

### C. Automated Program Repair

Huang et al. fine-tune code models ranging from millions to billions of parameters, including parameter-efficient methods, on bugs, vulnerabilities and compilation errors in Java, C and C++, and JavaScript, and show that fine-tuned models outperform earlier repair tools [6]. They also report that fine-tuning can erase knowledge gained in pre-training, that large models are costly to adapt, and that iterative repair can keep proposing the same wrong patch. In their setting every fix is generated. ProdPilot removes generation from the 44 rules whose fix is known in advance, and it bounds every run with three attempts per issue and five cycles, so a repair cannot repeat without limit.

Hao et al. improve pre-trained code models for repair by generating syntactically varied but equivalent training programs and presenting them in a curriculum from easy to hard, with gains over baselines on BFP and Defects4J benchmarks [7]. The experiments use Java only, the training schedule is fixed, and the defects are logic errors inside single functions. ProdPilot targets a different class of defect, operational configuration that spans several files of a project, and repairs it by template or extraction wherever the fix does not depend on the codebase's design.

### D. Static Analysis as Model Feedback

Dolcetti et al. ask open-source models to write C functions for 100 tasks, check the output with ground-truth tests and a static analyser, and then return the findings to the models for repair [8]. The models often produce incorrect or unsafe code, are poor at spotting their own errors, and repair well once given concrete feedback. The authors acknowledge that the tests may be incomplete and that the analyser can raise false positives. Their finding that models cannot judge their own code is the premise of ProdPilot's design: the agent's report of success is never used as the verdict, and the rule's checker decides every fix, including fixes that no model wrote.

Chapman et al. alternate calls to a static analyser for C with queries to a language model in order to infer which values each function returns on error [9]. On real systems code the combination lifts recall to 77.83%, against 52.55% for the analysis alone, while precision stays close to its previous level, and the authors study how the model's nondeterminism changes the analysis. The approach accepts that nondeterminism as part of the result. ProdPilot takes the opposite route where it can: values are extracted deterministically from the project, and when extraction is ambiguous the routine refuses and lists the candidates instead of asking a model to choose.

Alrashedy et al. return the findings of two static analysers to a language model so that it patches vulnerable Python code it has generated, and they release a benchmark of realistic security tasks for the purpose [10]. The method reduces detected vulnerabilities by up to 33%, but the authors note that flaws involving several functions or external systems remain and suggest runtime analysis and human involvement as future directions. ProdPilot adds the runtime evidence that static feedback cannot give, a real container build, a health probe and smoke tests against the live service, and it sends issues it cannot resolve to manual review with their exact location and reason.

Liao and Wang build a loop in which compiler messages, simulation logs and coverage results guide a model that writes hardware verification code, and they reach full functional coverage in a case study of one serial bus protocol [11]. They recommend treating every generated artifact as untrusted until an independent check has validated it. The evidence comes from a single case, and generation is used for every artifact the loop produces. ProdPilot applies the same untrusted-until-checked principle to all fixes, those from templates included, and evaluates it across 23 sample projects.

### *E. Constraining Agent Behaviour*

Darwish et al. constrain a model with rules written for the task, so that its answers in a dialogue-restructuring application stay within fixed limits, and they score the output by sequence matching against the source transcript [12]. They report an 85.5% improvement in the consistency of responses. The model still produces every answer, and the approach needs a validation scheme written for each task. ProdPilot takes the model out of the fixes that do not need it, and for the six that do, the validation scheme is the rule's own checker, which already exists.

Kazlaris et al. organise more than 300 studies of hallucination mitigation into six categories of method and point to open problems, including the lack of standard benchmarks and the difficulty of attributing gains when several methods are combined [13]. The survey compares methods for making generation more reliable but does not ask how much of a task can be done without generation. ProdPilot reports that share for its domain as a measured value, 88.0% of the ruleset.

Naser reviews the definitions, causes, tests and datasets for hallucination, the methods for detecting it, such as consistency checks, fact checking, calibration and retrieval, and the methods for correcting it [14]. The review calls for systems that can indicate the limits of their knowledge instead of filling gaps with invented content. In ProdPilot detection is deterministic, and an extraction routine that cannot decide says so by refusing and naming the candidates it found.

### *F. Container and Scanner Reliability*

Haq et al. show that container vulnerability scanners report different results for the same image and raise false positives, and they present LUCID, which stores the results of several scanners in a database and reconciles them by query [15]. LUCID reduces inconsistencies by 70% and predicts severity with 84% accuracy. It reconciles findings about an image that already exists but does not change the configuration that produced it. ProdPilot checks and repairs that configuration, the Dockerfile, the ignore files and the server setup, before the image is built, and it reports ambiguous scanner signals instead of acting on them.

Pietrikova´ et al. merge measures of size, complexity, rule violations and cloning into interpretable proficiency profiles and demonstrate them on student code written in C and Java [16]. The authors note that static measures are proxies, that the results depend on configuration and that their datasets are small. The profiles describe code but do not act on it. ProdPilot aggregates rule outcomes in a similar way, into a score and a feature vector, and uses both to drive repair and the release decision.

Lee et al. run three attacks on the networks linking Docker containers, namely ARP spoofing, denial of service and escalation of privilege, and analyse their effect on traffic, processor load and remote shell access [26]. The study concerns attacks on running containers rather than the way images are built. ProdPilot addresses the build side before deployment, for example by requiring the container to run as a user without root privileges and to be built in stages that leave build tools out of the final image.

### *G. Readiness Scoring and DevSecOps*

Lavin et al. define a scale of technology readiness for systems built on machine learning, a staged process in which a system advances through reviews and documentation, and illustrate it with case studies [21]. Readiness is assessed by reviewers against criteria; it is not predicted from outcomes. ProdPilot learns an estimate of deployability from 684 real deployments and uses it as one condition of an automated gate.

Prates and Pereira conduct a multivocal review of academic and industry sources to define DevSecOps practices and map them to supporting tools [22]. They list selection bias, the risk of outdated sources and inconsistent terminology in grey literature among their limitations, and the review does not measure how the practices affect deployment. ProdPilot encodes such practices as rules and measures, with the deployability model, which of them actually predict whether a project deploys.

Manolov et al. evaluate GitHub Advanced Security on GitHub and Azure DevOps, covering the scanning of code and of secrets and the analysis of dependencies inside delivery pipelines [23]. Their evaluation is qualitative and focused on design, without measurements of how quickly findings are fixed, and it assumes that people interpret and resolve the findings. ProdPilot links detection directly to repair and to independent verification, and it encrypts repository secrets against the repository's public key before sending them to GitHub.

Karamitsos et al. apply continuous integration and delivery practices to machine learning applications in order to shorten feedback loops, expose technical debt and simplify maintenance [20]. Their pipeline delivers models; it does not audit the application being delivered. ProdPilot audits, repairs and gates the application before its delivery pipeline is created.

### H. Delivery Pipeline Automation

Erdenebat et al. extend DevOps tooling so that many projects can have their own environments in a shared, self-hosted Kubernetes cluster, using customised Helm charts, and they validate the approach at a large software company [17]. The approach assumes that the projects entering the pipeline are already fit for production. ProdPilot places an audit, a repair loop and a gate in front of the pipeline, so that only a project that passes them is deployed.

Buttar et al. assemble a delivery pipeline from version control, Docker, Jenkins, cloud hosting and log monitoring for cloud applications and select resources by weighing cost against capacity, reporting lower cost than a benchmark solution [18]. The pipeline is designed by specialists for a given application. ProdPilot generates the container definition and the workflow for a vibe-coded project automatically and confirms that the workflow is active.

Bou Ghantous and Gill evaluate a DevOps reference architecture for deploying IoT applications to several clouds through an industry case study and a survey of experts [19]. The architecture guides the design of a pipeline but is not itself executable. ProdPilot is an executable pipeline whose final stage is checked against a live service.

Barbosa et al. use design science research to build a framework that combines DevOps practices with microservices, refine it through expert interviews and demonstrate it on one service [5]. They note a small sample and specific context, and they report that only about five percent of companies offer formal DevOps training. The framework assumes a team able to follow it. ProdPilot encodes the operational knowledge in rules and contracts, so that a developer without that training still has it applied.

### I. Summary of the Gap

Table I compares these groups of work. Each covers part of the path from detection to a verified deployment, and none covers all of it. The literature review in Table II and Table III lists the methodology and limitation of each study together with the way ProdPilot responds.

TABLE I
COVERAGE OF RELATED APPROACHES

| Approach | Aud. | Rep. | Ver. | Out. | Dep. |
|---|---|---|---|---|---|
| Scanners, static analysis [15], [16] | Yes | No | No | No | No |
| Program repair [6], [7] | No | Yes | P | No | No |
| Feedback loops [8]–[11] | Yes | Yes | P | No | No |
| Readiness, DevSecOps [21]–[23] | P | No | No | No | No |
| Pipeline automation [17]–[20] | No | No | No | No | P |
| ProdPilot | Yes | Yes | Yes | Yes | Yes |

Aud.: audits the project. Rep.: repairs it. Ver.: verifies every fix regardless of its author. Out.: predicts deployment from real outcomes. Dep.: deploys and verifies the live service. P: partly.

## IV. PROBLEM STATEMENT

Vibe-coded projects fail at deployment because nothing in them states what production requires, and nothing checks those requirements before a deploy is attempted. The research problem is narrower. Automated remediation of these requirements depends on free-form generation, which varies between runs and is seldom verified. We ask how much of that dependence can be removed by classifying fixes, and whether the remainder can be made safe by verification that does not depend on who wrote the fix. Table IV lists the design challenges and how each is met.

The first version supports Node.js 20 LTS with Express and React 18 with Vite 5, deploys to Render through a four-method provider interface, and freezes the ruleset at 50 rules so that its measurements stay comparable. VS Code with Copilot is the first target client, followed by Cursor and Windsurf.

## V. SYSTEM ARCHITECTURE

ProdPilot has five layers that run in order over one project directory (Fig. 1). Layer 0 identifies the stack and loads its blueprint, Layer 1 audits the project, Layer 2 repairs it, Layer 3 decides whether it may be deployed and Layer 4 deploys it. Everything runs on the developer's machine, no model inference takes place and no external AI service is called; network access is limited to deployment and to the package registry during the build check.

### A. Tool Interface

The IDE agent reaches ProdPilot through five MCP tools over stdio: a connectivity check, stack detection, a request for the fix contract of one failing rule, a report that a fix was applied, and deployment. ProdPilot only answers requests and never asks the client's model to generate anything, so a repair takes two calls: the agent asks for a contract, edits the files and reports back. The report is treated as a claim, and the answer comes from re-running the rule's checker. Four of the tools are marked as read-only for the client. The deployment tool is marked as changing its environment, and because it creates real infrastructure its description tells the agent to get the developer's approval before running it.

### B. Detection and Blueprints

Detection reads the file tree and `package.json` and needs neither a model nor the network. It refuses to guess: when one project declares both Express and React with Vite, it is reported as unrecognised. A blueprint lists what a deployable project of one stack must contain, 28 requirements for Express and 22 for React. The React blueprint marks the connectivity and API domains as not applicable, since a static site has no database connection and no API of its own, so the audit can tell an inapplicable domain from a forgotten one.

TABLE II
SUMMARY OF THE REVIEWED LITERATURE, PART I

| Ref. | Methodology | Limitation / research gap | How ProdPilot addresses it |
|---|---|---|---|
| [1] | Historical analysis of intent mediation; conceptual definition of vibe coding; research agenda | Names risks such as opaque code and unclear responsibility, but proposes no mechanism to contain them | Expresses deployability as 50 checkable rules, each with a verdict, a location and a named fix |
| [2] | Systematic review of 44 primary studies from 40,412 records, 2021 to 2025; three-domain risk taxonomy | Relies on human review; inconsistent human baselines across studies | Detects recurring flaws such as hardcoded credentials automatically and confirms each repair with the same check |
| [3] | Two models build one Spring Boot backend by prompt chaining; prompts, edits, requirements and tests measured | One application and two models; evaluation ends at code and tests, not deployment | Audits and repairs the operational layer, then proves it with a real build and deployment |
| [4] | Heuristic filtering of three million app reviews; 1,000 annotated; seven-type taxonomy | Observational; errors are found after release, by users | Moves detection before deployment and decides it with deterministic checkers |
| [24] | Open-source MCP server giving a model deterministic diagnostic tools; demonstration package | Not validated in the field; the model is trusted to choose tools and write conclusions | Never relies on a model's conclusion; validated with three real MCP clients |
| [25] | Survey of retrieval and MCP in education; proposed architecture; small proof of concept | Prototype without model generation, live MCP servers or human evaluation | Fully implemented server exercised end to end in VS Code, Cursor and Windsurf |
| [6] | Fine-tuning of code models of millions to billions of parameters, including parameter-efficient methods, on three defect types | Every fix is generated; costly adaptation; iterative repair can repeat wrong patches | 44 of 50 fixes need no generation; three attempts and five cycles bound every run |
| [7] | Code augmentation with curriculum learning to fine-tune pre-trained code models; BFP and Defects4J | Java only; single-function logic bugs; fixed training schedule | Repairs project-level operational defects across files, by template where possible |
| [8] | Open-source models write C for 100 tasks; tests and static analysis fed back for repair | Models cannot judge their own code; tests may be incomplete; analyser noise | The agent's report never decides; the rule's checker judges every fix |
| [9] | Static analysis interleaved with model queries to infer error specifications in C | Accepts the model's nondeterminism inside the analysis | Extracts values deterministically and refuses on ambiguity instead of asking a model |
| [10] | Static-analyser findings returned to a model to patch vulnerable Python; new benchmark | Static feedback only; flaws spanning several functions remain | Adds a real build and live smoke tests; unresolved issues go to review with location and reason |
| [11] | Loop with compiler, simulation and coverage feedback for generated verification code; one protocol | Single case study; generation used for every artifact | Checks template fixes as strictly as generated ones; evaluated on 23 projects |
| [12] | Task-specific rules bound a model's answers in a dialogue task; sequence-matching score | Output constrained but still generated; a validation scheme is needed per task | Removes the model from invariant fixes; the rule's checker is the validation scheme |

TABLE III
SUMMARY OF THE REVIEWED LITERATURE, PART II

| Ref. | Methodology | Limitation / research gap | How ProdPilot addresses it |
|---|---|---|---|
| [13] | Method-oriented taxonomy of over 300 hallucination-mitigation studies | No standard benchmarks; no measure of how much of a task can leave the model | Reports that share, 88.0% of the ruleset, as a computed value |
| [14] | Review of hallucination definitions, tests, datasets, detection and correction | Detection itself remains largely probabilistic | Detection is deterministic; extraction signals its limits by refusing with candidates |
| [15] | Database reconciliation of several container scanners; severity classifier | Reconciles findings about a built image; does not repair its configuration | Checks and repairs the configuration that produces the image |
| [16] | Static-analysis signals merged into interpretable proficiency profiles of C and Java code | Descriptive; proxy measures; small datasets | Aggregates rule outcomes into a score and features that drive repair and gating |
| [26] | Experimental network attacks on Docker containers; traffic, load and shell analysis | Addresses running containers, not how images are built | Enforces build-time hardening such as a non-root user before deployment |
| [21] | Staged readiness scale for systems built on machine learning, with reviews and case studies | Qualitative; assessed by reviewers rather than predicted from outcomes | Learns deployability from 684 real deployments |
| [22] | Multivocal review of DevSecOps practices and tools | A catalogue; selection and temporal bias; effect on deployment not measured | Encodes practices as rules and measures which of them predict deployment |
| [23] | Qualitative, design-oriented evaluation of a code-security suite on two platforms | No measured remediation; findings need human interpretation | Links detection to repair and verification; seals secrets before upload |
| [20] | CI/CD principles and tools applied to the delivery of machine learning models | Delivers models; the application itself is not audited | Audits, repairs and gates the application before its pipeline is wired |
| [17] | Per-project environments in shared Kubernetes clusters with customised Helm; industrial validation | Assumes production-ready projects; no audit or repair | Places an audit, a repair loop and a gate before deployment |
| [18] | Pipeline of version control, Docker, Jenkins, cloud hosting and log monitoring; cost-aware resources | Designed by specialists for one application | Generates and verifies the container definition and workflow automatically |
| [19] | DevOps reference architecture for multicloud IoT; case study and expert survey | Design guidance, not an executable tool | An executable pipeline checked against a live service |
| [5] | Design science framework combining DevOps and microservices; expert interviews; demonstration | Small sample; assumes a trained team, while about 5% of companies train formally | Encodes operational knowledge so that an untrained developer still has it applied |

### C. Audit Engine

Each blueprint requirement is enforced by one rule with an identifier, one of nine domains, a priority from P0 (critical) to P5, a scope, a check type and a fix type. Fig. 2 shows how the 50 rules spread over the domains and fix types. Of the

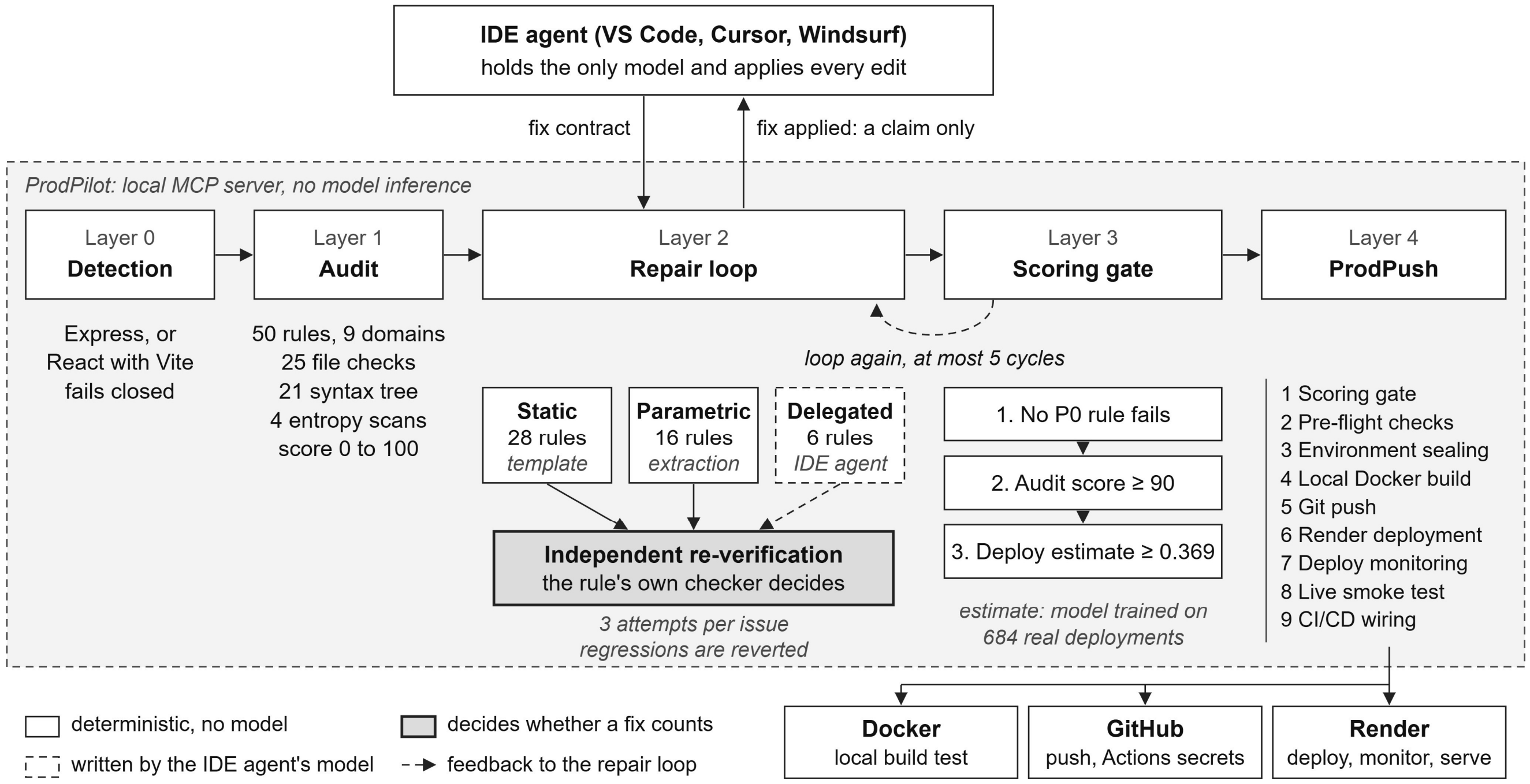

Fig. 1. Architecture of ProdPilot. The IDE agent, which holds the only model, applies every edit and exchanges fix contracts and reports with the server over MCP. In Layer 2 each fix is a template, an extracted value or an agent-written change inside a boundary (dashed), and every fix is accepted only by the rule's own checker. The gate evaluates its three conditions in order and returns a refused project to the loop for up to five cycles. Only Layer 4 reaches external services.

TABLE IV
DESIGN CHALLENGES AND MITIGATIONS

| ID | Challenge | Mitigation |
|---|---|---|
| C1 | Generated fixes vary between runs | Classify each rule before any fix exists; 44 of 50 never reach a model |
| C2 | Agent reports are unreliable | Accept a fix only if the rule's checker passes on disk |
| C3 | One fix can break another | Audit after each change and revert regressions |
| C4 | A good score can hide a critical failure | Refuse any project with a failing P0 rule |
| C5 | A model trained on compliance is circular | Label on real deployments and exclude the audit score |
| C6 | Repair may not terminate | Three attempts per issue and at most five cycles |
| C7 | Extraction can be ambiguous | Refuse and list the candidates instead of guessing |
| C8 | Build output is unbounded text | Sort failures into five fixed categories |
| C9 | Deployment handles live credentials | Keep credentials out of project trees and seal secrets |
| C10 | A successful deploy call is not a working service | Smoke-test the live service and confirm CI/CD is active |

50 rules, 25 read configuration files, 21 parse JavaScript into a syntax tree and 4 scan for secrets. Syntax trees are needed because order often matters more than presence: helmet must be registered before the routes it protects and the error handler after them, and only a tree preserves that order. Parsing uses acorn, vendored with its JSX plugin and run under Node, because the available Python port of esprima stops at ECMAScript 2017 and rejects constructs common in generated code, such as optional chaining and JSX fragments. Secret detection combines Shannon entropy with known key prefixes. Since no external call is allowed, a key cannot be tested against its provider, so borderline candidates are reported rather than dismissed, and committed history is scanned as well as the working tree.

The audit score weights each rule by priority, 32, 16, 8, 4, 2 and 1 from P0 to P5, and reports the share of applicable weight earned by passing rules. A project that fails one P0 rule and nothing else still scores 93 on the Express ruleset and 91 on React, so the score alone cannot keep critical failures out; the gate treats them separately (Section V-E).

### D. Fix Classification and the Repair Loop

Every rule carries one of three fix types, and the rule schema rejects a rule whose fix reference does not match its type. *Static* fixes are identical for every project and come from a template. *Parametric* fixes have a fixed shape filled with values read from the project, such as the Node version and start command of a Dockerfile; when a value is ambiguous, for instance with two lockfiles present, the routine refuses and lists the candidates. *Delegated* fixes depend on how a codebase is organised and are written by the IDE agent inside a boundary. Table V gives the split. Four of the six delegated rules concern code structure; the other two concern secrets in committed history, where the right repair depends on how far a secret has spread.

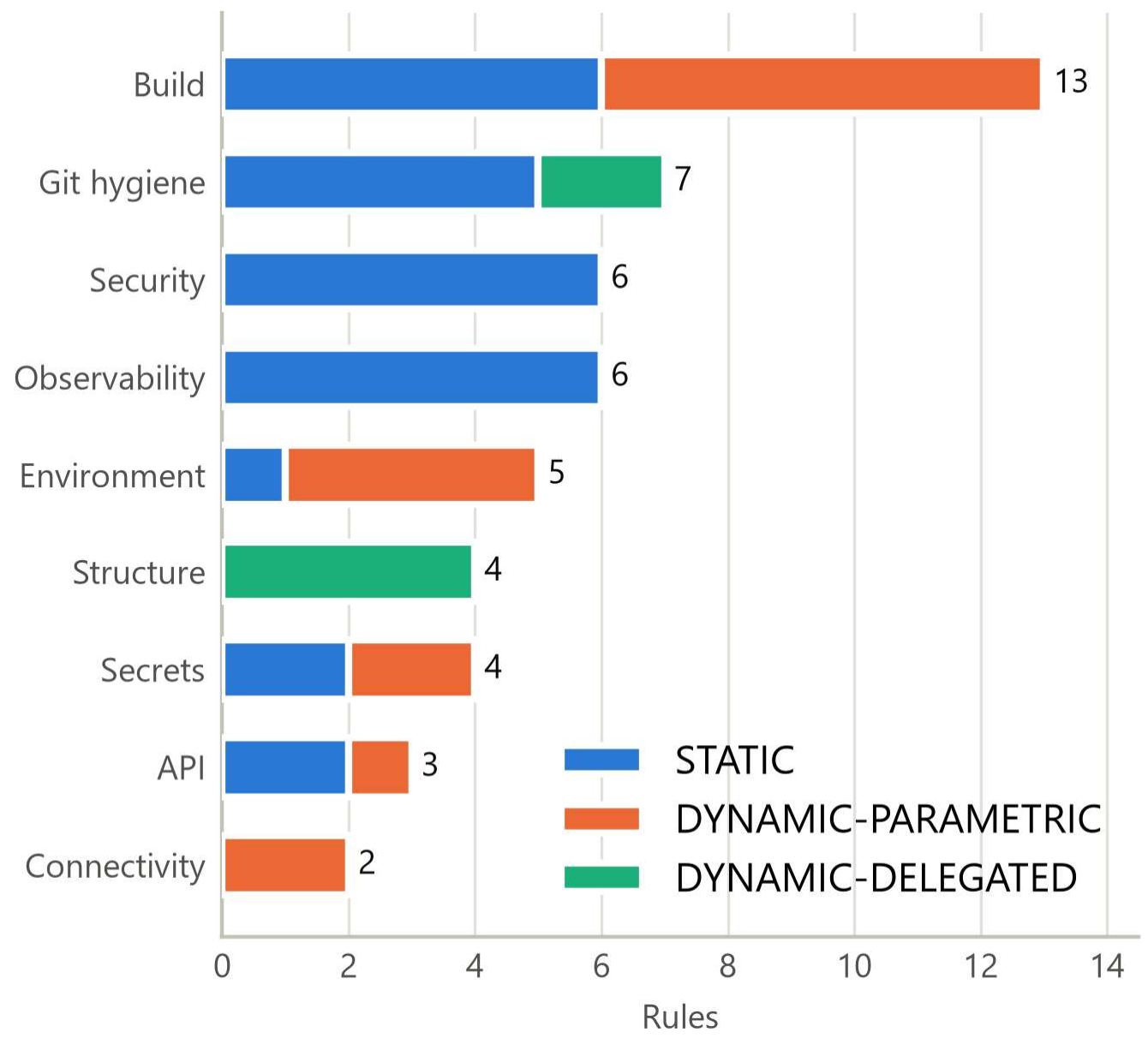

Fig. 2. The 50 rules of the frozen ruleset by domain, split by fix type. The structure domain is delegated in full, and two Git hygiene rules concerning secrets in committed history are delegated; every other domain is repaired by template or extraction.

TABLE V
FIX TYPES OVER THE RULESET

| Stack | Static | Param. | Deleg. | Total | No model |
|---|---|---|---|---|---|
| Node.js, Express | 16 | 9 | 3 | 28 | 89.3% |
| React, Vite | 12 | 7 | 3 | 22 | 86.4% |
| Combined | 28 | 16 | 6 | 50 | 88.0% |

A contract for a static or parametric fix carries the exact content, an anchor naming where it goes, the packages and environment keys the content needs, and an instruction to change nothing else. A delegated contract carries the violation, the requirement written as a condition the checker evaluates, the files that may change and what must stay unchanged, and it names the function that will judge the result. The requirement given to the agent is therefore the same predicate used to accept its work.

The loop takes failing rules one at a time in priority order and re-runs the rule's checker after each change. An issue has three attempts over the whole run, one that cannot even be attempted goes straight to manual review, and a run is limited to five cycles. After any change that alters a file, a full audit looks for regressions; if a rule that passed before now fails, the change is reverted and the issue goes to manual review. This guard was added after the first end-to-end run exposed defects that isolated tests had missed (Section VII-D).

### E. Scoring Gate

When a cycle ends, the gate audits the whole project again instead of trusting the loop's record, because a fix can break a rule that was never queued. It then applies three conditions in order: no P0 rule may fail, the audit score must be at least 90, where the Production Ready band begins, and the deployability estimate must reach the model's operating point of 0.369. A project that fails any condition returns to the loop, within the five-cycle limit. The model is consulted last, so it may veto a project the rules accept but can never admit one they reject, and a missing input such as an unreadable model or an undetermined build keeps the gate closed. An earlier design that replaced the audit score with the model's probability at the same threshold was measured and rejected, since no project in the dataset reaches 0.9 even with every failing rule cleared.

### F. Deployability Model

The model estimates the probability that a project will go live on Render and answer its health path. Its 26 features are the numbers of rules assessed and failed in each of the nine domains, the failures at each of the six priorities, the stack, and whether the project compiles under the build command Render itself uses, run locally in a Linux container. The audit score is excluded because it is a fixed function of those counts, and the labels come from real deployments rather than from the rules, so the model cannot simply reproduce the audit.

The estimator is a histogram gradient boosting classifier with a monotonic constraint: more failures can only lower the estimate and a successful build can only raise it, so a repair never makes a project look less deployable. Training checks this on every row after calibration and rejects any model that violates it. Probabilities are calibrated with a sigmoid over five folds, and the operating point is the cut with the highest F1 on out-of-fold predictions.

### G. Deployment Pipeline

ProdPush runs the gate and then eight stages: checking prerequisites, sealing the environment, building and probing a local container, pushing the generated files, creating the Render service, monitoring the deploy, smoke-testing the live service and wiring CI/CD. It stops at the first failing stage and names it. Credentials are stored under the user's home directory, outside every project; on Windows an owner-only file mode left the credential file readable by five other principals, so access is set through the access control list and read back before it is reported. Configuration values reach the service only through the Render API, and repository secrets are encrypted as libsodium sealed boxes before they are sent to GitHub. Build and deploy failures are sorted into five fixed categories, so only structured issues can reach the repair loop. Monitoring maps Render's eleven deploy states onto three, and the smoke test retries the health path with backoff for up to 90 seconds before checking an API route, the security headers, the CORS header and the absence of stack traces. The pipeline deploys first and wires CI/CD second, because a workflow cannot run until the service and its identifier exist;

the workflow triggers deploys through the Render API, since a deploy hook cannot be read programmatically, and waits for the deploy it started before checking health. The provider interface was checked against a second implementation that shares nothing with Render but its four method names.

## VI. Implementation Details

This section describes how ProdPilot was built and tested, as distinct from what it does.

### *A. Technology Stack and Code Base*

ProdPilot is a Python package, released as version 1.0.0rc1, of 43 modules and about 15,200 lines of source code, and it supports Python versions 3.11 through 3.14. Table VI lists the main technologies. scikit-learn is fixed at the exact release used to save the model, because a saved estimator is only guaranteed to load under that version, and the trained model ships inside the package so that an installed copy needs nothing else to run the gate.

TABLE VI
Implementation Technologies

| Component | Technology | Used for |
|---|---|---|
| Protocol server | MCP Python SDK 2.0 | Five tools over stdio |
| Command line | Typer | serve, setup, doctor, connect |
| Code parsing | acorn, acorn-jsx, Node | Syntax trees for 21 rules |
| Containers | Docker SDK for Python | Build test, build feature |
| Hosting | Render REST API | Services, values, states |
| Repository | Git, GitHub REST API | Push, secrets, Actions |
| Secret sealing | PyNaCl (libsodium) | Repository secrets |
| Model | scikit-learn 1.9.0, joblib | Deployability estimate |

### *B. Client Integration*

The command `prodpilot connect` writes the configuration each client needs and refers to the installed command by its full path. For VS Code it uses VS Code's own command-line interface, writing into the user profile, and the result is read back from VS Code's configuration; Cursor is configured through its global settings file; Devin, the current name of Windsurf, is configured through a file local to the project, because it blanks the workspace variable that the other clients' files use. That local file names a path valid on one machine, so the command warns if Git would commit it. The command `prodpilot doctor` reports on the stored credentials, on whether any other account can read their file, and on the presence of Git, Node.js and a running Docker daemon.

### *C. Rules, Contracts and the Loop*

The rule store is data rather than code: each rule is one entry, declared in the same order as the blueprint it enforces, and a test fails if the two sets drift apart. Static templates place their content through ten named anchors, such as the position that follows creation of the Express app and precedes the first route, because a literal line number cannot be promised across projects. Parametric routines return either a value or a refusal that lists the candidates, and they read the project again on every attempt, since an earlier fix may have created the file a later rule needs. On the seeded project the regression guard costs roughly 3.6 seconds per full loop.

### *D. Data Collection, Labelling and Training*

Repositories were found with GitHub's search API, kept within its rate limits. The verdict on each was written down before moving to the next, which lets a stopped collection continue from where it stopped. Only a manifest of name, commit and licence is stored, so the dataset can be rebuilt at the same commits. Each labelled project was deployed to Render on the free plan, React projects as static sites and Express projects as web services, with service creation paced to the platform's limit of 20 per hour; every service was deleted as soon as it was labelled, and a final query confirmed that none was left running. The build feature was computed by running Render's build command in a Linux container on a read-only copy of each project, with the Node version Render would select. Training used a stratified split with a fixed seed and five-fold cross-validation over a grid of 108 parameter combinations. The chosen configuration has trees of 7 leaves with at least 20 rows in each, 200 boosting iterations at a learning rate of 0.05 and an L2 penalty of 1.0. The model was promoted only after a second training run reproduced its predictions exactly on every row.

### *E. Deployment Stages*

Stage 3 gives the container the project's own port setting, otherwise the port its image exposes, otherwise Render's default of 10000, tries each published port in turn until one responds; if the container exits at start-up, the stage records the exit code together with the output line that gives the cause. Stage 4 commits nothing but ProdPilot's own generated files, will not push while the developer has other uncommitted work, and supplies the GitHub token inside the push address for that single command, never saving it in the repository's configuration. Monitoring polls every 15 seconds with a timeout of 10 minutes, and the smoke test retries the health path after 1, 2, 4, 8 and 15 seconds within its 90-second window.

### *F. Portability to Other Providers*

The pipeline reaches Render through a protocol of four methods: one starts a deployment (`deploy`), one reports what became of it (`poll_status`), one returns its output (`get_logs`) and one writes configuration values (`set_env`). The protocol was checked against a second implementation that shares nothing with the Render client but those four names. Everything above that protocol is provider-agnostic: the audit, the rules and their contracts, verification and the regression guard, the loop and the gate's audit conditions, the local container build, and the generation of the Dockerfile and the workflow. What is specific to Render is narrow and named: the mapping of its eleven deploy states onto three, the build command used to compute the build feature, the choice between a static site and a web service, the default port of 10000, and the polling interval and timeout. What is specific to GitHub is the sealing of repository secrets against the repository's public key, the Actions workflow, and triggering a deploy through the provider's API rather than a deploy hook. Porting therefore means writing those four methods and one state mapping; the stages themselves do not change.

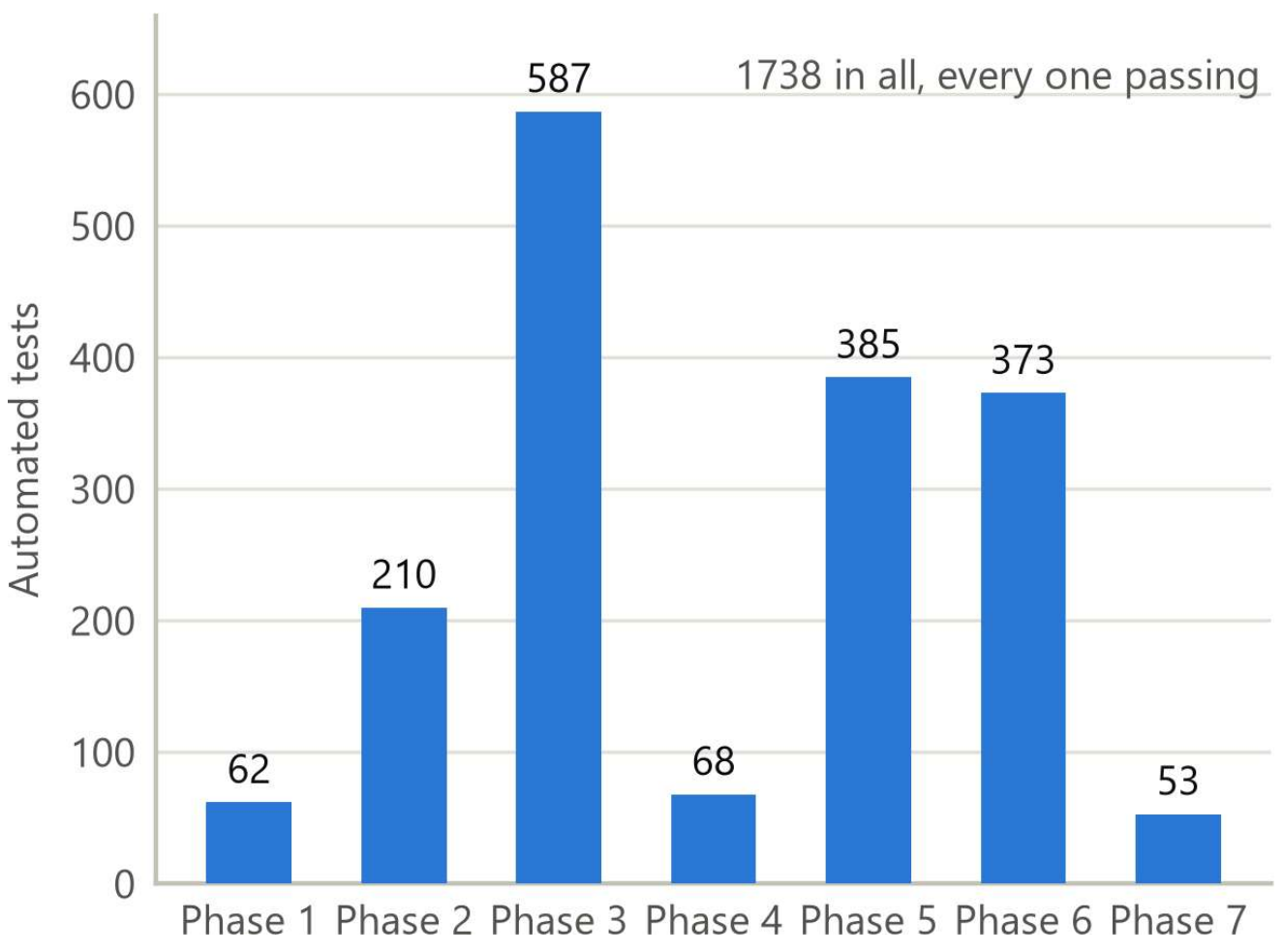

Fig. 3. Automated tests written in each implementation phase, 1,738 in total at the close of the seventh phase, all passing.

The deployability model is the part that does not travel. Its labels are real Render outcomes and its strongest feature is defined by the build command Render runs, so an estimate learned here does not describe another platform, and a new provider would need its own labelled set before the gate's third condition could be used. The first two conditions, no failing critical rule and a score of at least 90, are properties of the project rather than of the platform and transfer unchanged, so a port without a model is a gate with two conditions rather than none.

### G. Testing

Automated tests were written with each phase (Fig. 3); the suite held 1,738 tests at the close of the seventh phase, and the release candidate holds 1,796, which pass on Python 3.11 and 3.14 with a Docker daemon available. No test reaches a live service. Render and GitHub are replaced by scripted responses in the shapes their documentation describes, pushes go to real bare Git repositories on disk, and container tests build real images when a Docker daemon is present. The suite starts the server with the exact command each committed client configuration names, so a configuration that stops working fails the suite. Rule behaviour is exercised on 22 committed sample projects covering clean, broken, ambiguous and unsupported cases on both stacks.

### H. Experimental Environment

The experiments ran on Windows 11 with Python 3.11.15 and Docker. The client observations used VS Code 1.137.0 with GitHub Copilot Chat 0.65.0, Cursor 3.20.17 and Devin 3.10.23, and were recorded on 14 September 2026. Labelling and the live deployment used Render's free plan, and CI/CD ran on GitHub Actions.

## VII. Evaluation

### A. Setup

Two sets of inputs are used. The first is the corpus of 22 committed sample projects plus one derived variant, which gives a clean React sample the usual Vite index page and a lockfile so that it can be built. A harness ran every sample through the complete chain, from audit, loop and gate to the nine ProdPush stages, and recorded each stage's outcome. Everything except external services was real: the audit, the contracts and the file edits they produced, verification and the regression guard, the gate together with the promoted model and a Docker build of the project, and the code of all nine stages. Each sample was a genuine Git repository; stage 3 built the image and started the container, and stage 4 committed and pushed to a bare repository on local disk. The Render and GitHub APIs, and the deployed application that the smoke test probes, were replaced by scripted answers, since no run was allowed to reach a live service. A stand-in executor took the place of the IDE agent and applied content contracts literally. A delegated contract deliberately contains no content, and the executor cannot compose one, so all delegated rules in this run came back unapplied.

The second input is the dataset for the deployability model (Fig. 4, Table VII). Of 2114 public repositories found through the GitHub API, 872 had no licence, 40 had a non-permissive licence and 702 were not a supported stack; 500 were kept, and one could not be parsed. We added 186 synthetic negatives, each a working repository with one further rule deliberately broken, drawn from 25 rules and checked with the verifier to break nothing else. These are controlled perturbations, not models of how projects fail: a repository in this corpus already fails 14.8 rules on average, between 6 and 23, not one of the 499 fails a single rule alone, and every one of them fails at least one critical rule, 4.6 on average. Each of the 684 projects was deployed to Render over about 35 hours. A project is labelled positive if its service became live and its health endpoint responded, which held for 129 (18.9%). The stricter label of passing all five smoke checks could not be used, because no project passed it, mostly for lack of security and CORS headers. Nine projects whose build could not be determined were excluded, leaving 675 rows, split into 506 for training and 169 held out.

### B. Determinism and Fix Reliability

The design-time share of 88.0% in Table V is computed by the rule store, and every rule has a working fix path. The full-chain run adds a measure of the share at run time over the sample corpus: of the 269 rule instances for which the loop sent at least one contract, 201 were static and 59 parametric, so 260 (96.7%) needed no model-written content.

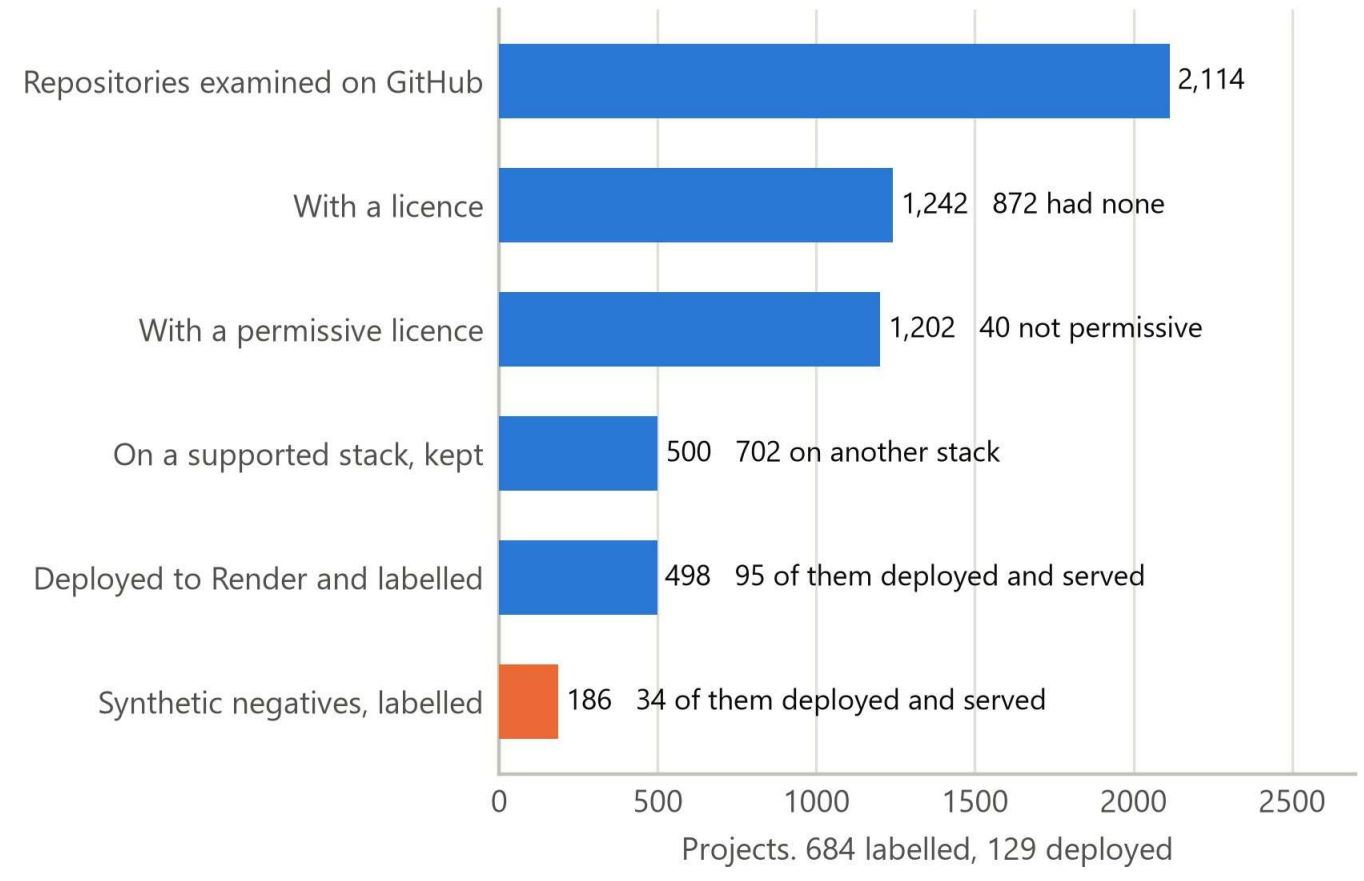

Fig. 4. Construction of the labelled dataset, from the repositories examined on GitHub to the projects deployed to Render and labelled, including the synthetic negatives.

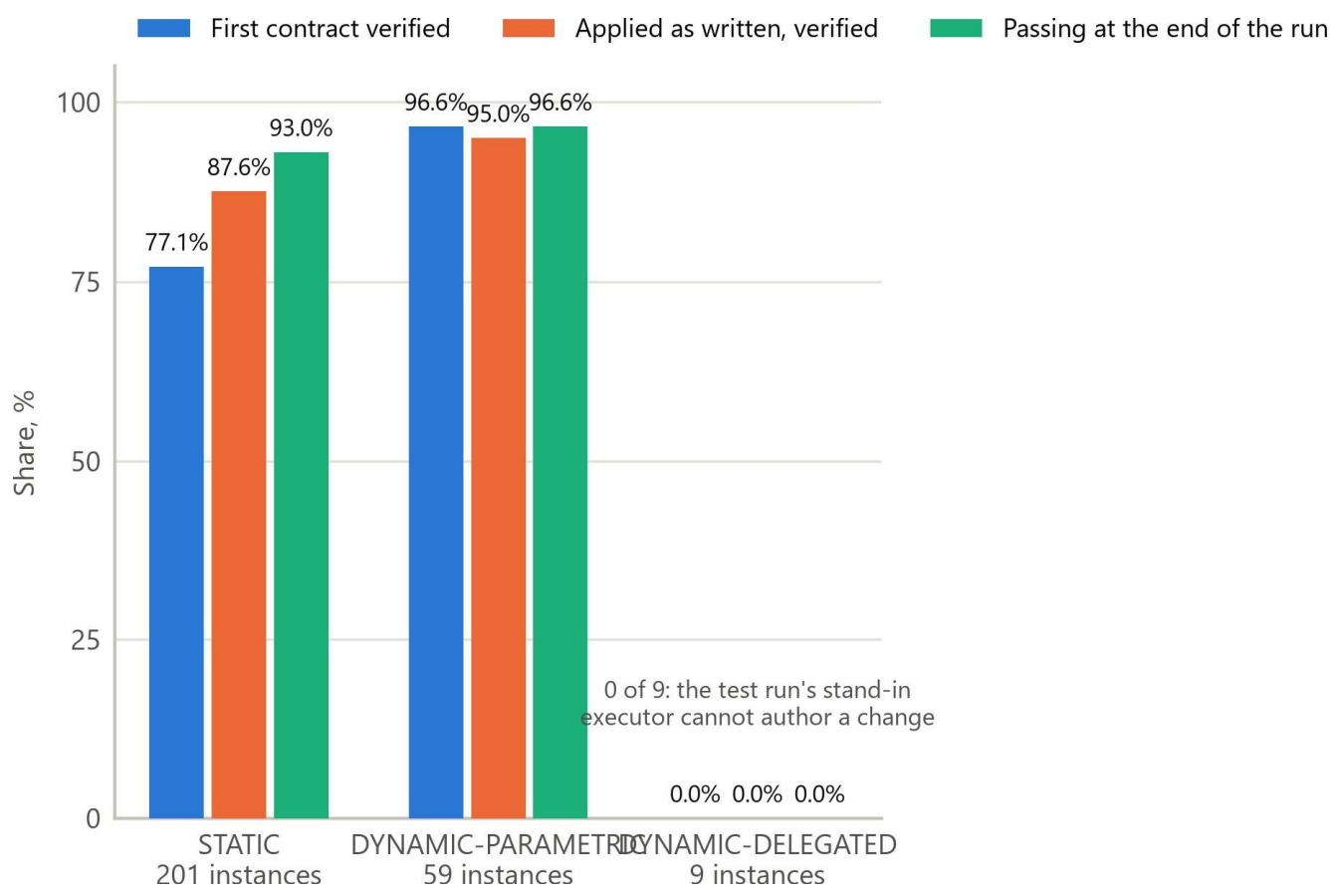

Fig. 5. Fix reliability by fix type over the full-chain run of 23 samples. The delegated rate is zero because the stand-in executor cannot write a change; it measures the executor, not an agent.

TABLE VII
LABELLED DEPLOYMENT DATASET

| Source | Projects | Deployed |
|---|---|---|
| Real repositories, Express | 259 | 8 |
| Real repositories, React | 239 | 87 |
| Synthetic negatives, Express | 68 | 12 |
| Synthetic negatives, React | 118 | 22 |
| Total | 684 | 129 (18.9%) |

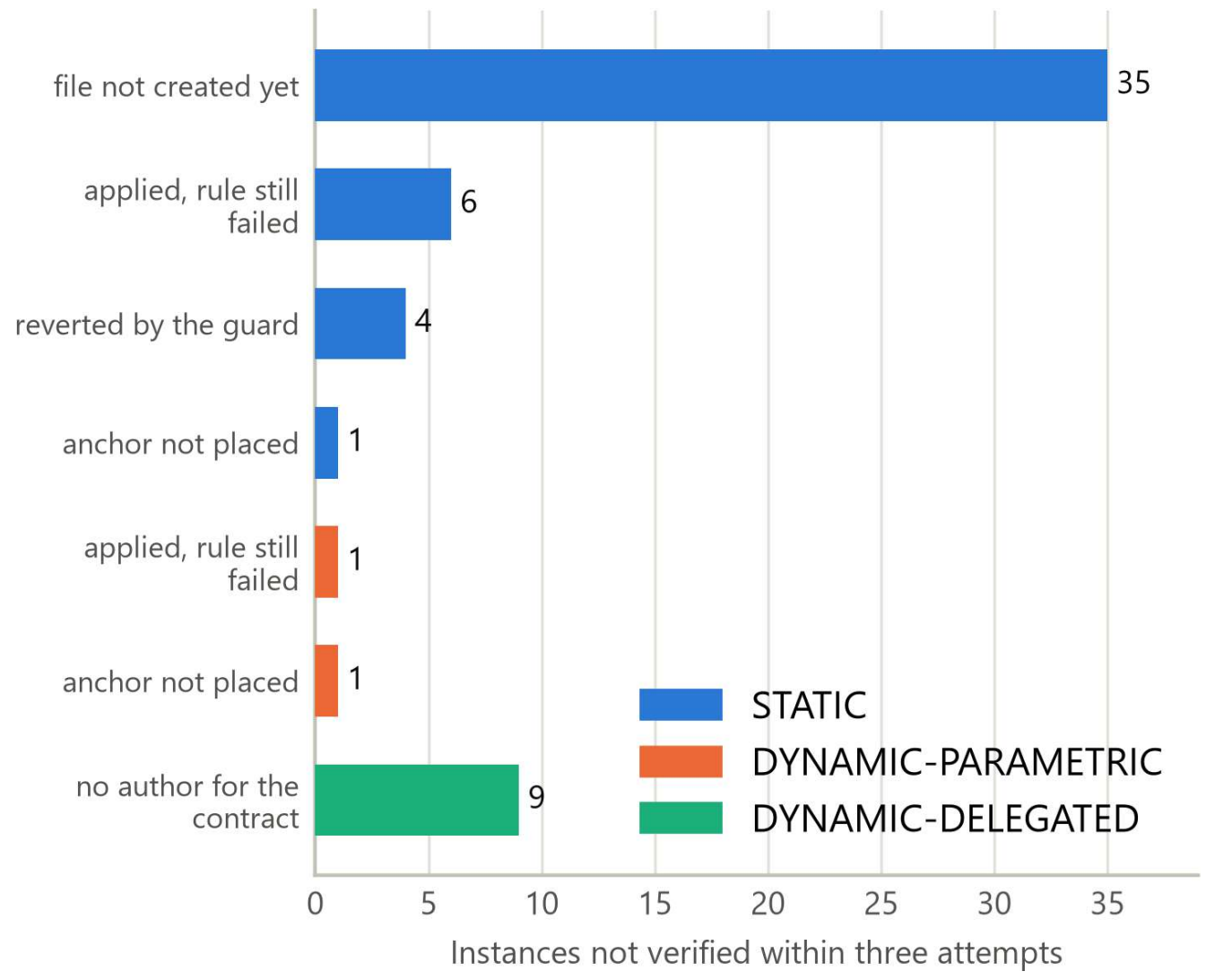

Fig. 6. Reasons for which rule instances failed verification within the three-attempt budget, by fix type.

Those are planted defects. To answer RQ2 over violations nobody planted, we audited the 499 processable repositories of the collected corpus, real projects taken from GitHub at pinned commits. They yield 7407 violations: 4881 (65.9%) carry a static fix, 2045 (27.6%) a parametric fix and 481 (6.5%) a delegated one. The deterministic classes thus cover 6926 violations, 93.5% of what a real corpus presents, above their 88.0% share of the ruleset because the rules that recur most often are those whose fix is fixed or extractable. The share is 94.9% within Express and 91.3% within React. Delegated work does not vanish: 412 of the 499 projects (82.6%) carry at least one delegated violation, so most projects still need the agent for something, but that is about one violation per project against a mean of 14.8. All 50 rules fire at least once in the corpus.

Fig. 5 reports three measures of reliability per fix type. Of the 177 static contracts the executor applied as written, 155 passed their checker (87.6%); the first contract succeeded for 77.1% of static rule instances, and 93.0% of them were passing when the run ended. Of the 60 parametric contracts applied, 57 passed (95.0%). A static contract always produces the same change for the same project, so it is deterministic, but it is not always sufficient. Fig. 6 gives the reasons. The largest group, 35 instances, consists of critical rules whose target file is created only later in the run, by a rule of lower priority; each passes once that file exists, unless the rule that creates it was itself refused. The remaining shortfalls are a sample with three entry points, in which each contract repaired its named file while the rule kept failing on a different one, an error handler that needed relocating, which an insertion cannot do, a file holding four credentials from which each attempt removed only one before the budget ran out, and fixes reverted by the regression guard. In no case was the content placed in the named file incorrect. After a later change that makes content contracts declare the environment keys they read, the four reverted fixes no longer occur and the static rate of applied contracts that pass rises to 89.8%.

In 15 of 74 parametric rule instances (20.3%) extraction declined to proceed and no contract was issued; each refusal listed the competing candidates (Fig. 7). The samples that trigger these refusals were built to be ambiguous, with two lockfiles, two database drivers, three entry points or routes under two roots, so the rate describes the corpus as much as the routines. Nine instances of the six delegated rules received 27 constraint contracts and none passed, which reflects the executor rather than an agent. Delegated fixes are measured separately, with a live agent, in Section VII-C.

### C. Delegated Fixes with a Live Agent

The rates above measure the stand-in executor, which cannot author a change at all. The delegated class itself was therefore measured with live agents, on 16 September 2026, over ten trials covering all six delegated rules (Table VIII). Two clients were used: Cursor 3.20.17 in agent mode driving Grok 4.6, and a Claude Code session driving Claude Opus 5 at high reasoning effort over the same stdio transport. In each trial the agent was given the project path and the identifier of the failing rule and asked to use the tools. Nothing told it what to write, and the outcome was decided by the rule's own checker rather than by the agent's report.

Every trial was resolved on the first attempt. For the four structural rules the agents moved a database query into a service module, made a route handler delegate to its controller, wrapped a React root in an error boundary they wrote, and added a catch-all route with the component it renders. For the three hygiene rules, where the fix is to remove a credential from committed history, the agents rewrote the commit that carried it; an inspection of the repository afterwards found the credential in neither the history nor the working tree, and in each case a second rule that had flagged the same credential in the source cleared as well. Across the ten trials every file changed lay inside the contract's boundary, a full audit found no rule that had been passing and now failed, and the agent's report agreed with the checker every time. Where the same rule was given to both clients, the two agreed exactly, including the files they chose to change.

Three limits bound this result. It is one trial per rule on one day, which establishes that the delegated path works rather than a rate at which it works. The eight Claude Code trials called the tools directly over stdio, so they exercise the contract rather than an IDE's approval loop, which the Cursor trials do cover. And for the hygiene rules the checker sees only the repository: rotating the exposed credential happens at the provider, outside the boundary the contract defines, and it was not done, so those three fixes are complete in the repository and incomplete operationally.

### D. Repair and the Gate

The first end-to-end run on a real project exposed defects that no isolated test had caught. Applied together and retried, the fix contracts inserted content repeatedly and required packages they never declared, and one inserted a metrics route above the security middleware, breaking a rule that had passed. The audit score rose, but the project no longer started. The contracts were made idempotent and made to declare their dependencies, and the regression guard was added.

After these corrections, one loop cycle on a deliberately broken Express project raised its score from 13 to 88 and resolved 17 rules, and the full gate took it to 96 (Table IX). The five issues left for manual review are ones an insertion cannot fix: two rules whose attempts ran out before the file they depend on was created, an error handler that has to be relocated rather than inserted, and the two delegated structural rules, which no automated test can write. A small Express API used for the deployment in Section VII-H went from 69 to 100 within a single cycle, introduced no regressions and still served every route.

Fig. 8 shows the audit score of every scored sample before the loop and at the gate. Eight of the 23 samples cleared the gate. For seven, at least one P0 rule was still failing, although two of these had reached a score of 90. The model alone turned away five React samples scoring 99 or 100, giving each a 5 to 6% chance, since none contains the index page from which Vite builds; the variant with that page received 83% and was admitted. The last three were not on a supported stack and stopped at the gate. Each of the three conditions of the gate

TABLE VIII
DELEGATED FIXES AUTHORED BY LIVE AGENTS

| Rule | Client | Att. | Files changed | Regr. |
|---|---|---|---|---|
| STR-001 | Cursor | 1 | Controller, new service | None |
| STR-001 | Claude Code | 1 | Controller, new service | None |
| STR-002 | Cursor | 1 | Route file | None |
| STR-002 | Claude Code | 1 | Route file | None |
| STR-003 | Claude Code | 1 | Root file, new boundary | None |
| STR-003* | Claude Code | 1 | Root file, new boundary | None |
| STR-004 | Claude Code | 1 | Router, new component | None |
| GIT-003 | Claude Code | 1 | Commit rewritten | None |
| GIT-003* | Claude Code | 1 | Commit rewritten | None |
| GIT-007 | Claude Code | 1 | Commit rewritten | None |

16 September 2026. Cursor 3.20.17 drove Grok 4.6; the Claude Code session drove Claude Opus 5 at high reasoning effort over stdio. Att.: attempts until the rule's own checker passed, of a budget of three. Regr.: rules passing before the change that failed after it. Every file changed lay inside the contract's boundary. Starred rows are a second project exercising the same rule: STR-003 on a project already scoring 99, and GIT-003 on a project with no entry point.

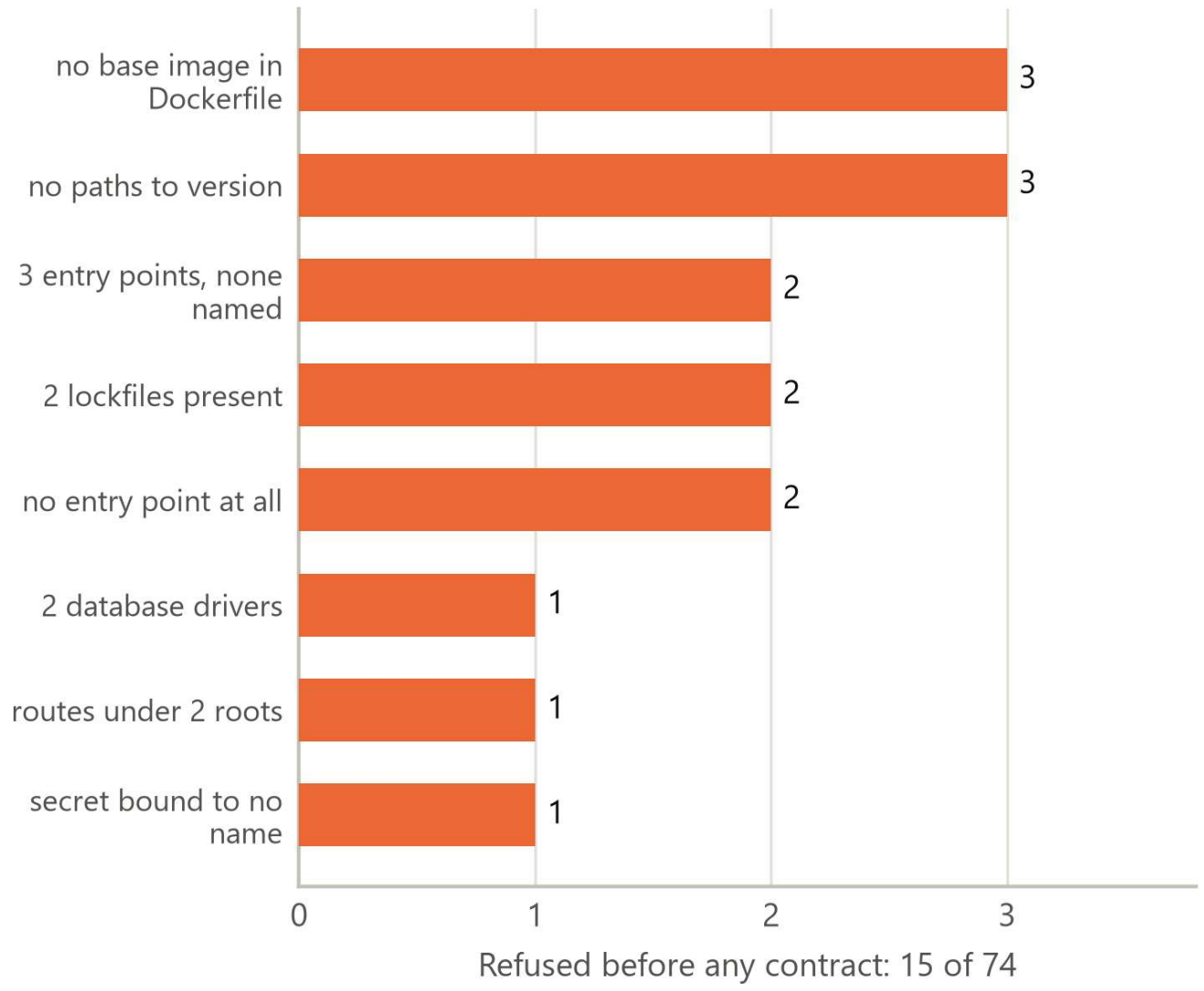

Fig. 7. Parametric rule instances that extraction declined, grouped by the reason each refusal gave.

TABLE IX
SEEDED PROJECT THROUGH THE LOOP AND GATE

| Measure | Result |
|---|---|
| Audit score | 13 before, 96 after |
| P0 failures | 6 before, 0 after |
| Cycles used | 2 of 5 |
| Left for manual review | 5, each with location and reason |
| Gate decision | Ready to deploy |

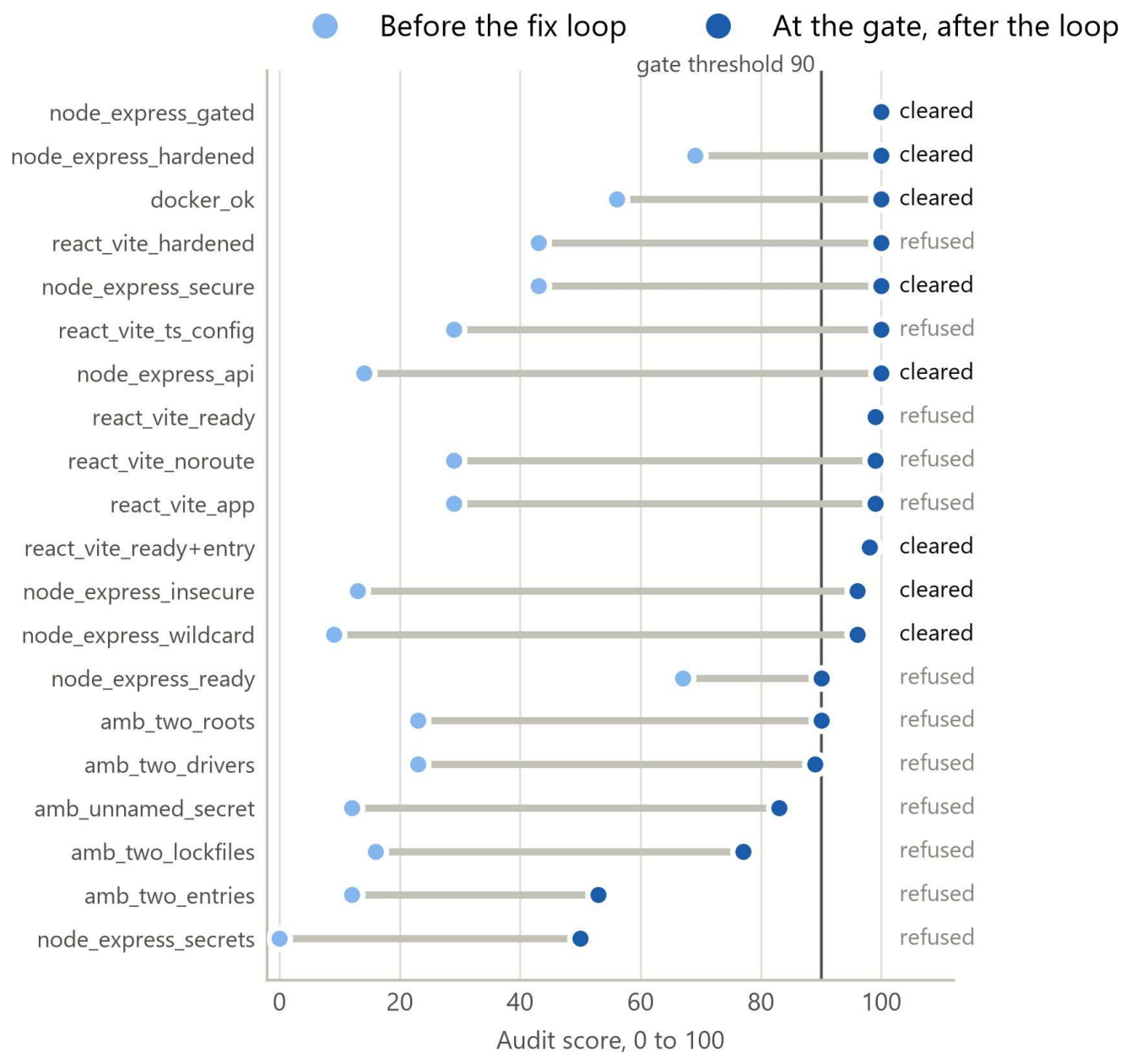

Fig. 8. Audit score of each sample before repair and on reaching the gate, with the gate's decision. A score of 90 or more is necessary but not sufficient: a failing P0 rule or a low deployability estimate also refuses a project.

therefore decided cases that the other two did not.

## E. Verification Independence

Table X reports the design-level evidence for RQ3. The checker that decides each outcome is the audit's own function, so the loop and a full audit cannot disagree; across five sample projects they agreed on every rule. In the decisive test the security middleware is registered after the routes, the exact mistake the rule targets: the agent reports success and the issue stays open.

TABLE X
VERIFICATION COVERAGE

| Property | Result |
|---|---|
| Rules shown failing on a violating project | 50 of 50 |
| Rules shown passing on a compliant project | 50 of 50 |
| Rules judged by the audit's own checker | 50 of 50 |
| Agreement with a full audit, five projects | Every rule |
| False report of success rejected | All three fix types |
| False report of failure rejected | All three fix types |

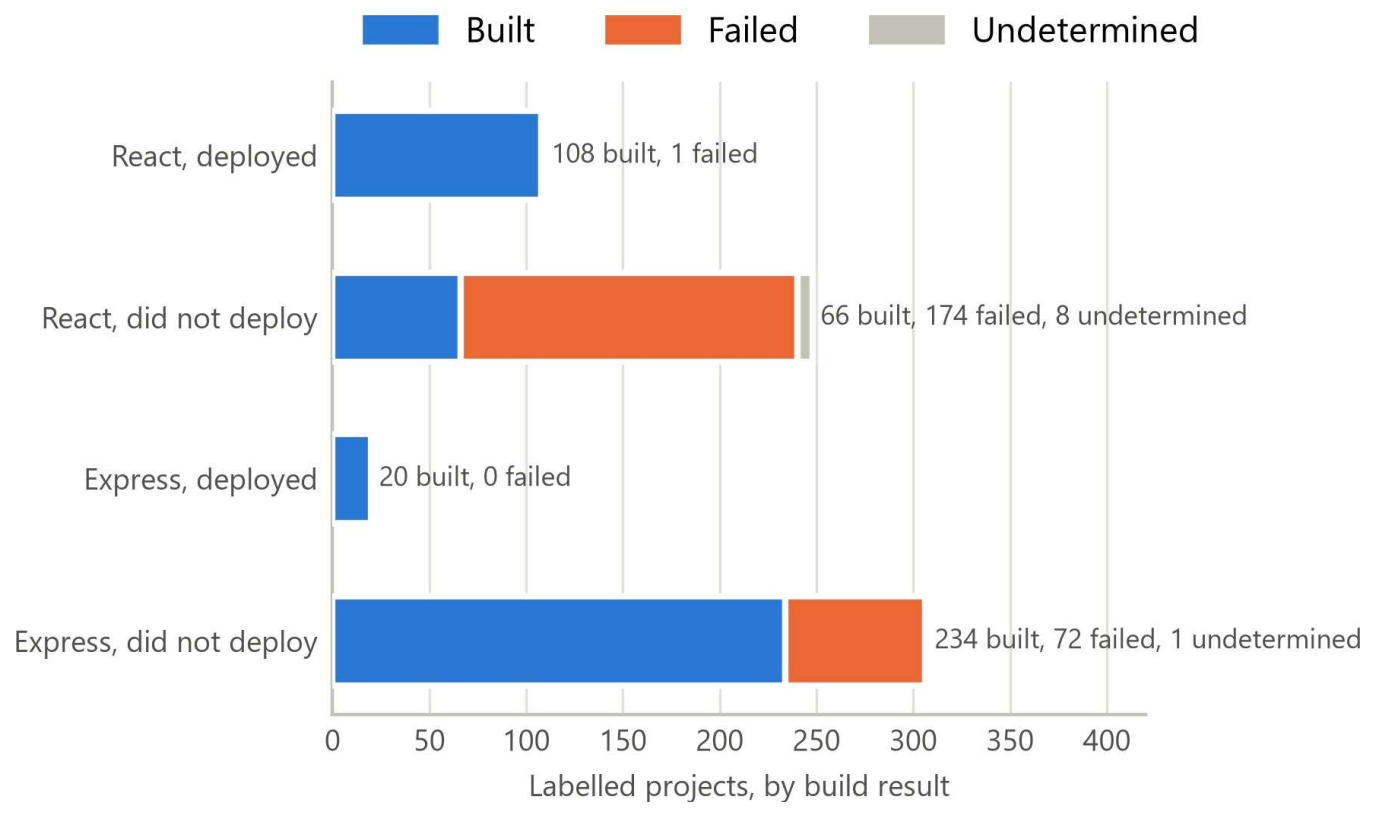

Fig. 9. Build results of the labelled projects under Render's own build command, by stack and by real deployment outcome.

The full-chain run gives a count. The executor reported 237 static and parametric contracts as applied, and verification confirmed 212 of them. Had the loop accepted the executor's reports, 25 fixes (10.5%) would have been recorded as successful: after 21 of them the rule still failed, and the other four passed their own rule but broke one that had passed before, so the regression guard reverted them. The reporting party in this run is the stand-in executor rather than a language model, and a live agent could err more often or less; what the run establishes is that none of these 25 cases passed independent verification.

## F. Deployability Model

The build result is the feature that separates the outcomes (Fig. 9). Among React projects, 108 of the 109 that deployed also built, while 174 of those that did not deploy failed to build; among Express projects the build rarely failed, and deployment depended on other factors.

We trained five models on one training split to separate the effect of the build feature from that of the monotonic constraint (Fig. 10). Models with the build feature reach a held-out ROC AUC above 0.90, and those without it stay below 0.78. The constraint costs no measurable ranking: on the training folds the constrained model D reaches a cross-validated PR AUC of 0.731 against 0.713 for the unconstrained model E, so E's higher held-out PR AUC reflects one split with 32 positives. Without the constraint, 2393 of 7281 single-rule fixes lowered the estimate of model C and 3234 lowered that of E; under D none did, so model D was promoted.

On the held-out rows model D scores 0.909 ROC AUC and 0.565 PR AUC (Table XI, Fig. 11, Fig. 12); a baseline that predicts from the stack only reaches 0.269 PR AUC, and one that always predicts failure 0.189. At the operating point the model finds 28 of the 32 held-out deployers with a precision of 0.583. In every calibration bin holding more than five rows, predicted and observed deployment rates differ by less than six percentage points (Fig. 13). The Express figures rest on five held-out deployers and should be read with that in mind.

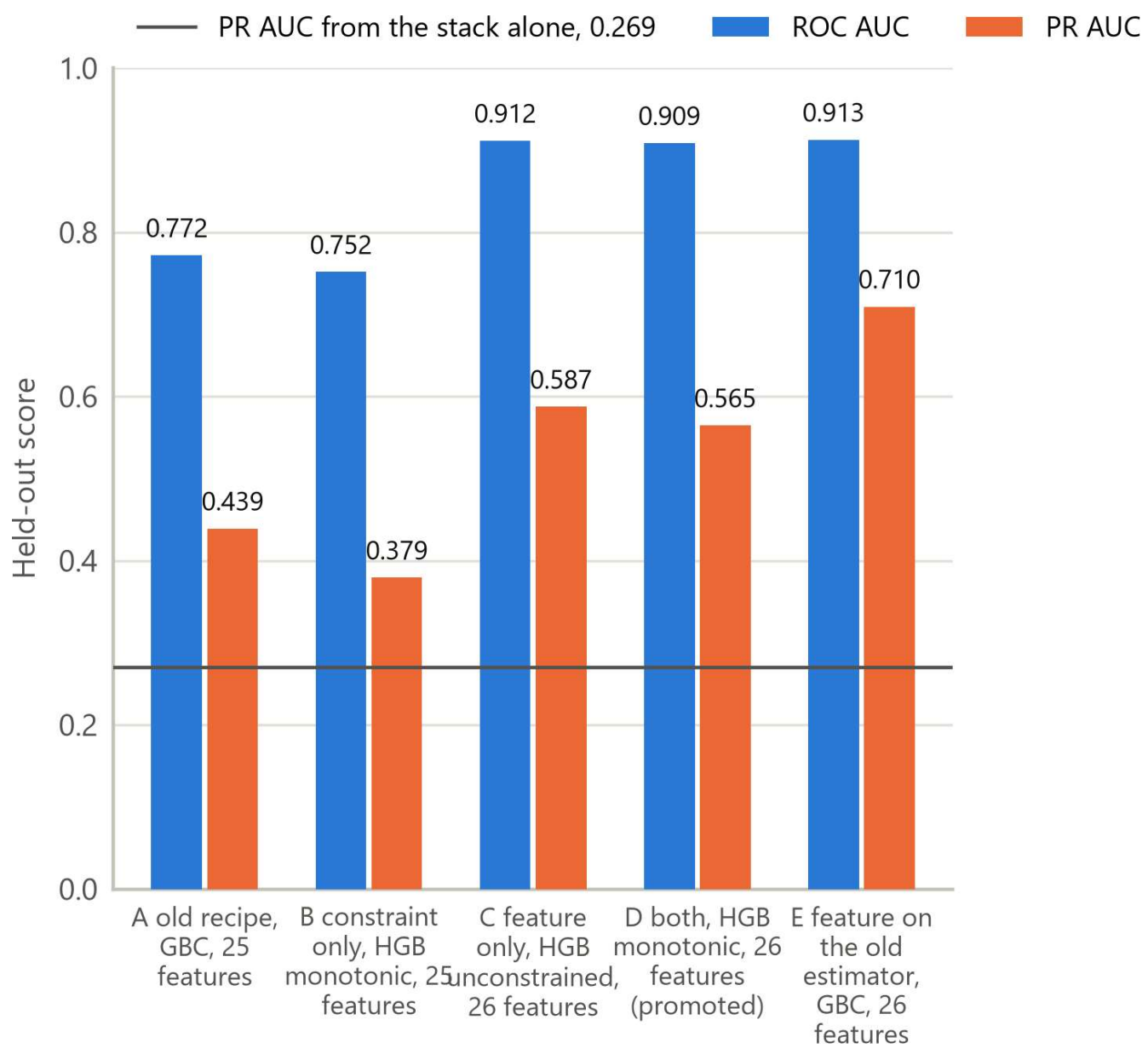

Fig. 10. Held-out ROC AUC and PR AUC of five models trained on the same split, against the PR AUC of predicting from the stack alone. Model D, histogram gradient boosting with the build feature and the monotonic constraint, is the model in use.

TABLE XI
MODEL D ON THE HELD-OUT ROWS

| Measure | Both | React | Express |
|---|---|---|---|
| Rows (deployed) | 169 (32) | 95 (27) | 74 (5) |
| ROC AUC | 0.909 | 0.852 | 0.913 |
| PR AUC | 0.565 | 0.567 | 0.566 |
| Brier score | 0.091 | 0.126 | 0.047 |
| Precision at 0.369 | 0.583 | 0.587 | 0.500 |
| Recall at 0.369 | 0.875 | 1.000 | 0.200 |

Permutation importance attributes most of the ranking to the build result, whose shuffling lowers the held-out PR AUC by about 0.31, while the next feature lowers it by 0.051 and the rest by less than 0.01 (Fig. 14). The estimate is therefore mostly about whether a project builds.

That concentration deserves a direct answer, since a model that leans on one deterministic check invites the question of what the other 25 features earn. The ablation in Fig. 10 answers it: without the build feature, models A and B still reach held-out ROC AUC of 0.772 and 0.752 and PR AUC of 0.439 and 0.379, against 0.269 for predicting from the stack alone and 0.189 for the prevalence. Audit counts alone therefore carry real signal, and the build result supplies the rest. Part of the concentration is mechanical. For a React project the platform serves the output of a build rather than a running server, so a successful build is close to the decision the platform itself makes, which is also why React separates so well; on Express the build rarely fails and other factors decide. The design consequence is the important one. Because a single feature dominates, the gate never delegates its decision to the model: the two audit conditions are evaluated first and on asks for is still open.

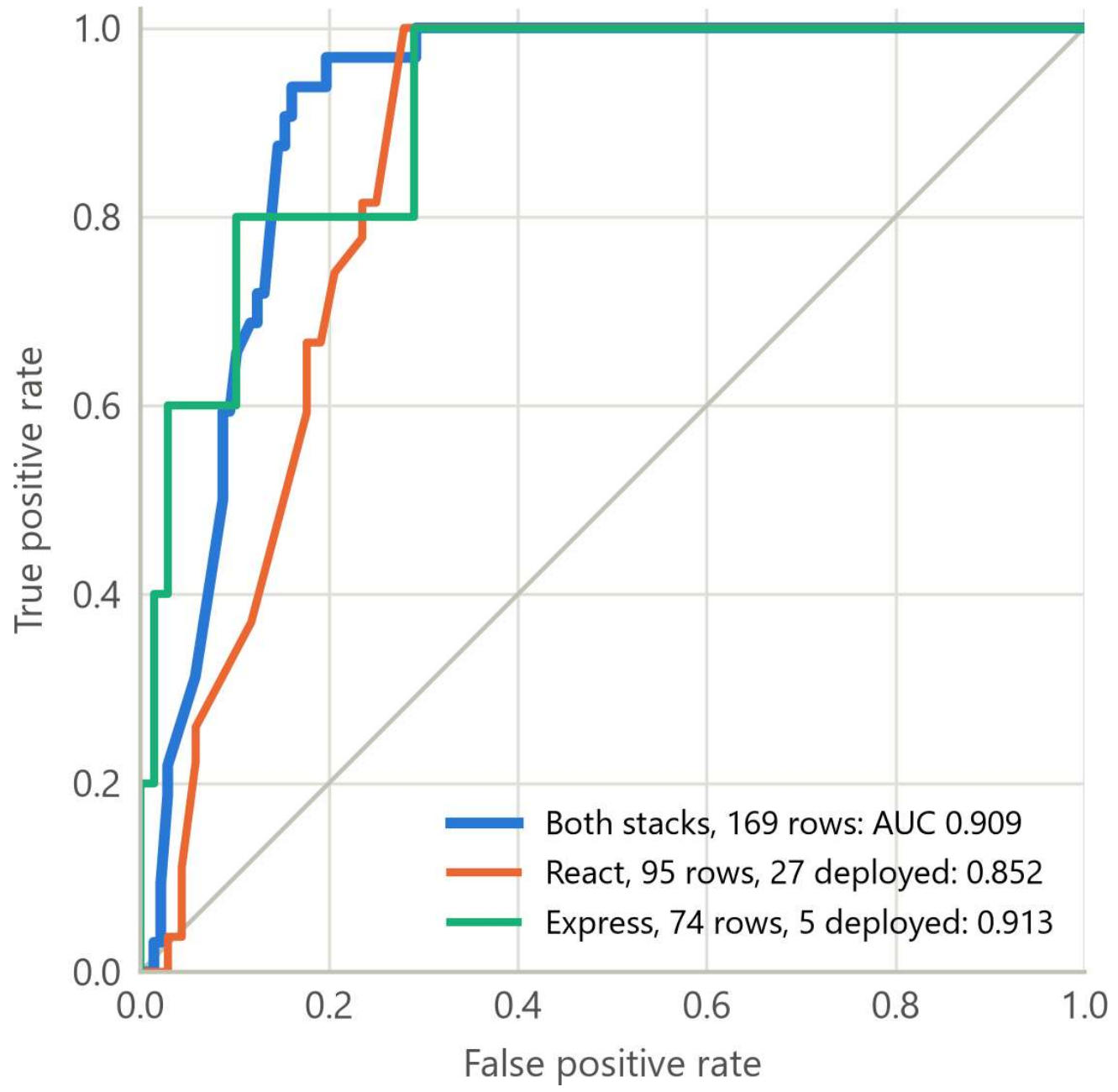

Fig. 11. ROC curves of model D on the 169 held-out rows, for both stacks and for each stack alone.

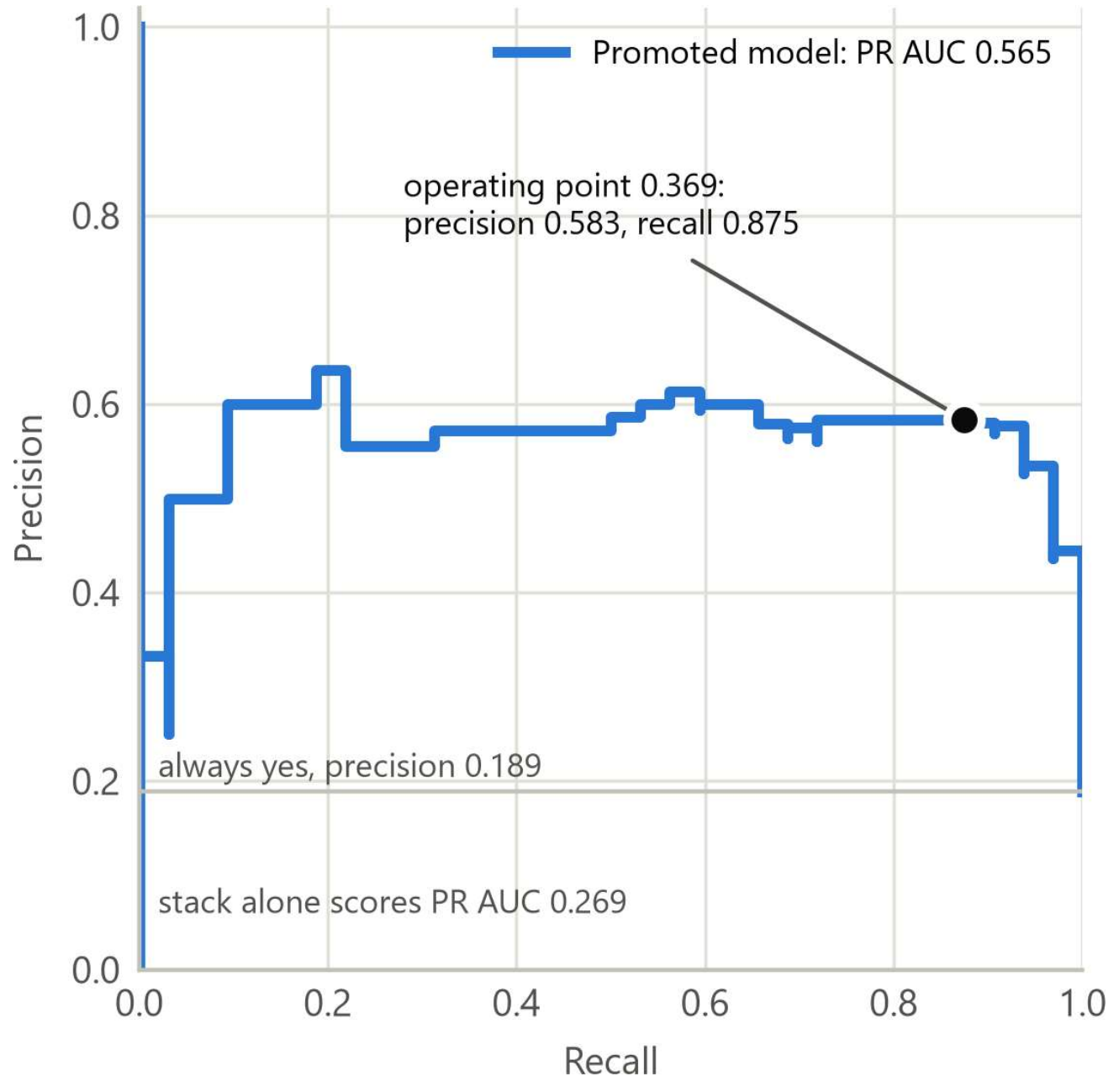

Fig. 12. Precision-recall curve of model D on the held-out rows, with the operating point and the two baselines.

The gate policy for React was decided against a rule fixed before the promoted model was evaluated. With every failing rule cleared, the earlier model gave React projects that deployed and those that did not the same mean estimate, 0.200 against 0.201, a ROC AUC of 0.489. The rule demanded that fully fixed React deployers be separable from non-deployers at 0.70 ROC AUC or more, both over all rows and over held-out rows, that the held-out ROC AUC within React also reach 0.70, and that three quarters of fixed deployers clear the operating point. Model D reached 0.864, 0.830, 0.852 and 108 of 109, so the model gates both stacks on equal terms. Table XII shows the effect on projects built for real: the fixed React project that cannot build is refused at 0.047, while an otherwise identical copy that builds is admitted at 0.851.

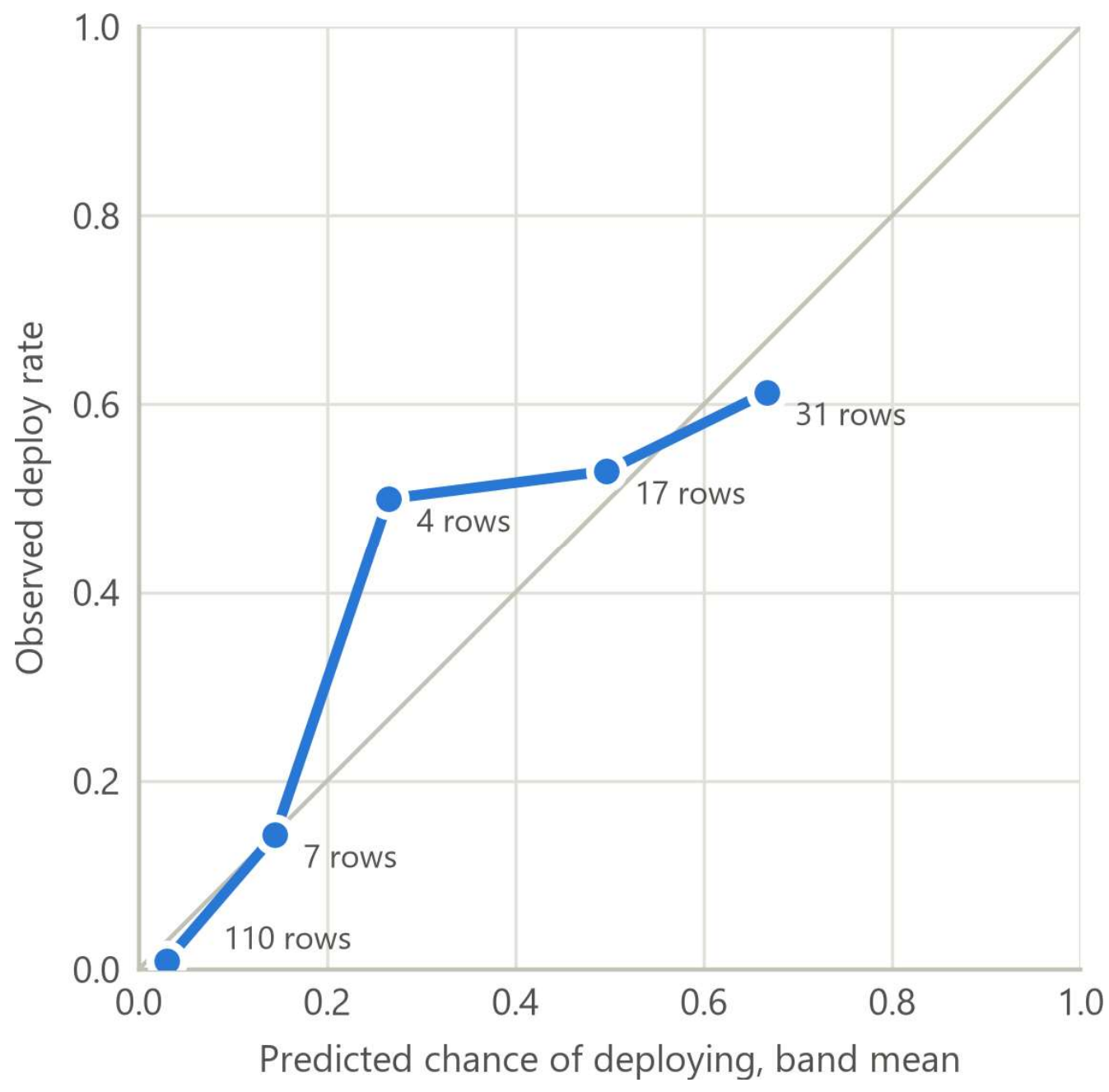

Fig. 13. Calibration of model D on the held-out rows: mean predicted chance of deploying against the observed deployment rate in each band, with the number of rows per band.

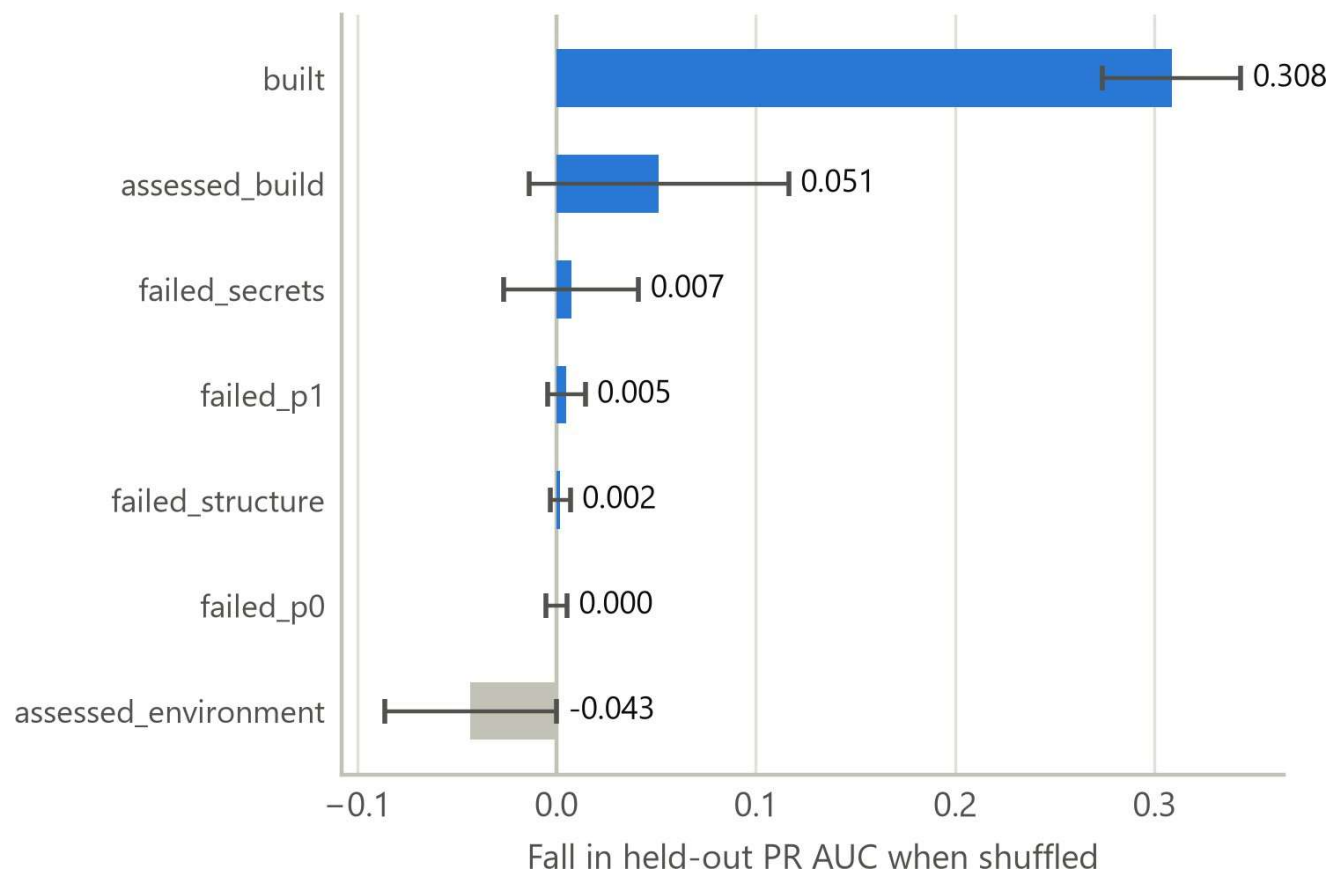

Fig. 14. Permutation importance of the features of model D, measured on the held-out rows as the drop in PR AUC after one feature's values are permuted, with the mean and spread over 30 repetitions.

their own, and the estimate can only veto a project they accept, never admit one they reject (Section V-E). The estimate is best read as a learned statement about buildability, with a smaller contribution from the audit, and it costs a real container build to obtain, which is the most expensive step of the gate.

Two observations bear on RQ4. The synthetic negatives, working repositories with one rule deliberately broken, deployed at 18.3% against 19.1% for real repositories, and no project passed all five smoke checks. Rule compliance and deployment outcome are thus different quantities, which is why labels come from deployments. A model trained on compliance labels was not built, so the direct comparison RQ4

TABLE XII
GATE DECISIONS WITH REAL BUILDS

| Project | Score | P0 | Est. | Decision |
|---|---|---|---|---|
| Express, fixed | 100 | 0 | 0.842 | Opens |
| React, fixed, no build | 99 | 0 | 0.047 | Refused, model |
| React, fixed, builds | 99 | 0 | 0.851 | Opens |
| React, partly fixed | 43 | 3 | 0.705 | Refused, P0 |
| React, unfixed | 28 | 3 | 0.595 | Refused, P0 |

### G. Full-Chain Integration

Before integration, the chain did not complete for any sample. Five broken Express projects were repaired to scores of 96 to 100, yet all five, together with the clean demonstration project and the React variant, failed when stage 3 built and ran them in Docker. Six defects in completed phases were behind this. Stage 3 gave the container no port setting and could reach only a port declared in the image; the generated Dockerfiles ran `npm ci` in projects without a lockfile; a critical rule failed a port read into a variable before use; ProdPilot's own nginx file failed its logging rule because it turns logging off for the health path alone; a repository public key that could not be decoded made the pipeline raise an error instead of stopping cleanly; and ProdPilot's React Dockerfile ran nginx without write access to its cache, so each React container it repaired exited at once. Each was corrected with its own tests.

With these corrections in place, the gate admitted 8 of the 23 samples and 6 finished every stage (Fig. 15). Four of those six had begun below 90 and were raised past it by the loop, from 14, 13, 9 and 56 to scores of 96 or 100. The two that cleared the gate and then stopped at stage 3 are sample projects that require modules they do not contain, and stage 3 now reports the application's own error. No run recorded an unhandled failure. A second run after the contracts began declaring their environment keys raised the gate count to 11 and the number of completed chains to 7, with no fix reverted by the regression guard.

### H. Deployment on Real Infrastructure

The Express API repaired from 69 to 100 was deployed through ProdPush against Render and GitHub in a single call.

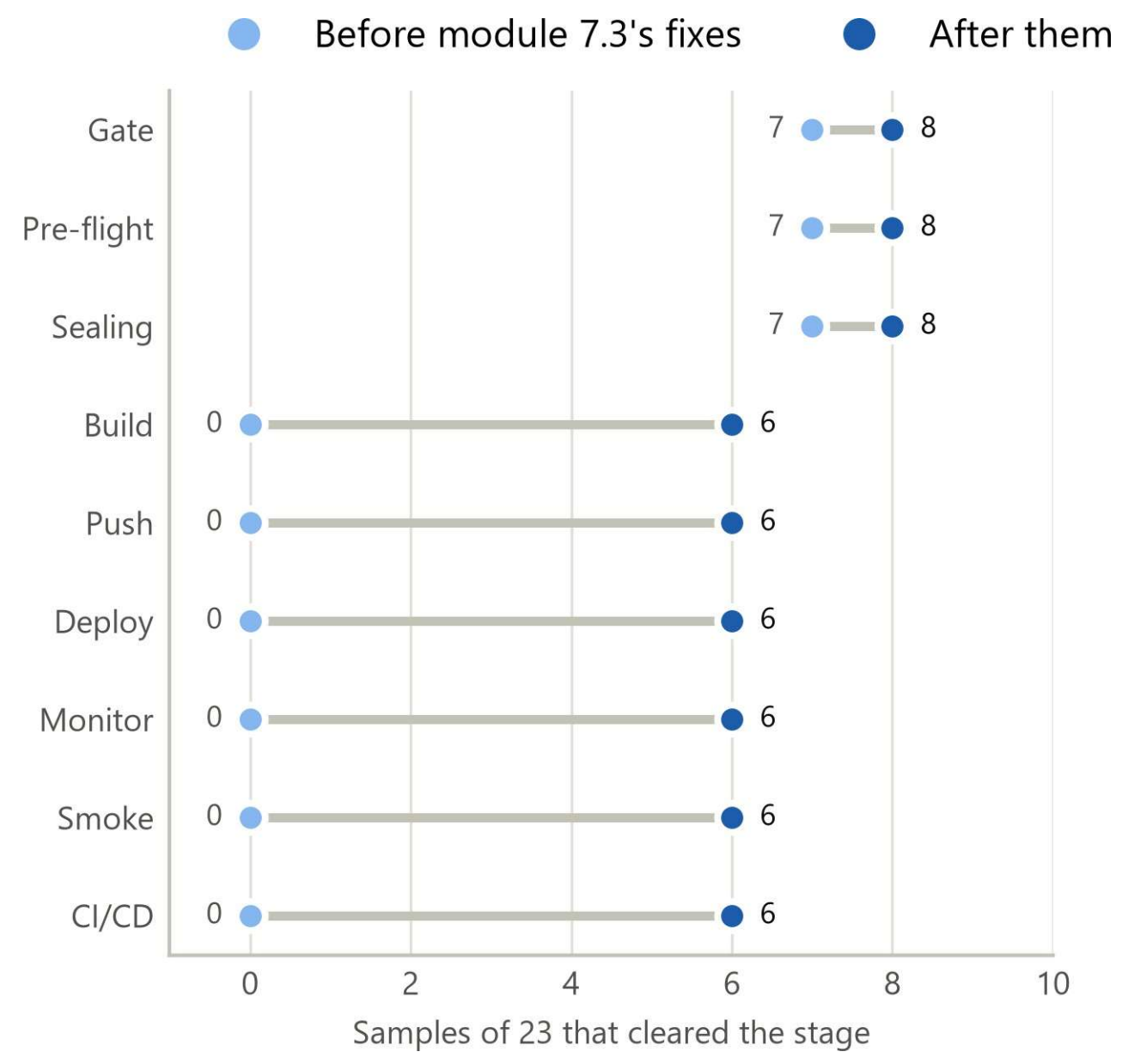

Fig. 15. Number of the 23 samples that cleared each ProdPush stage before and after the corrections made during integration.

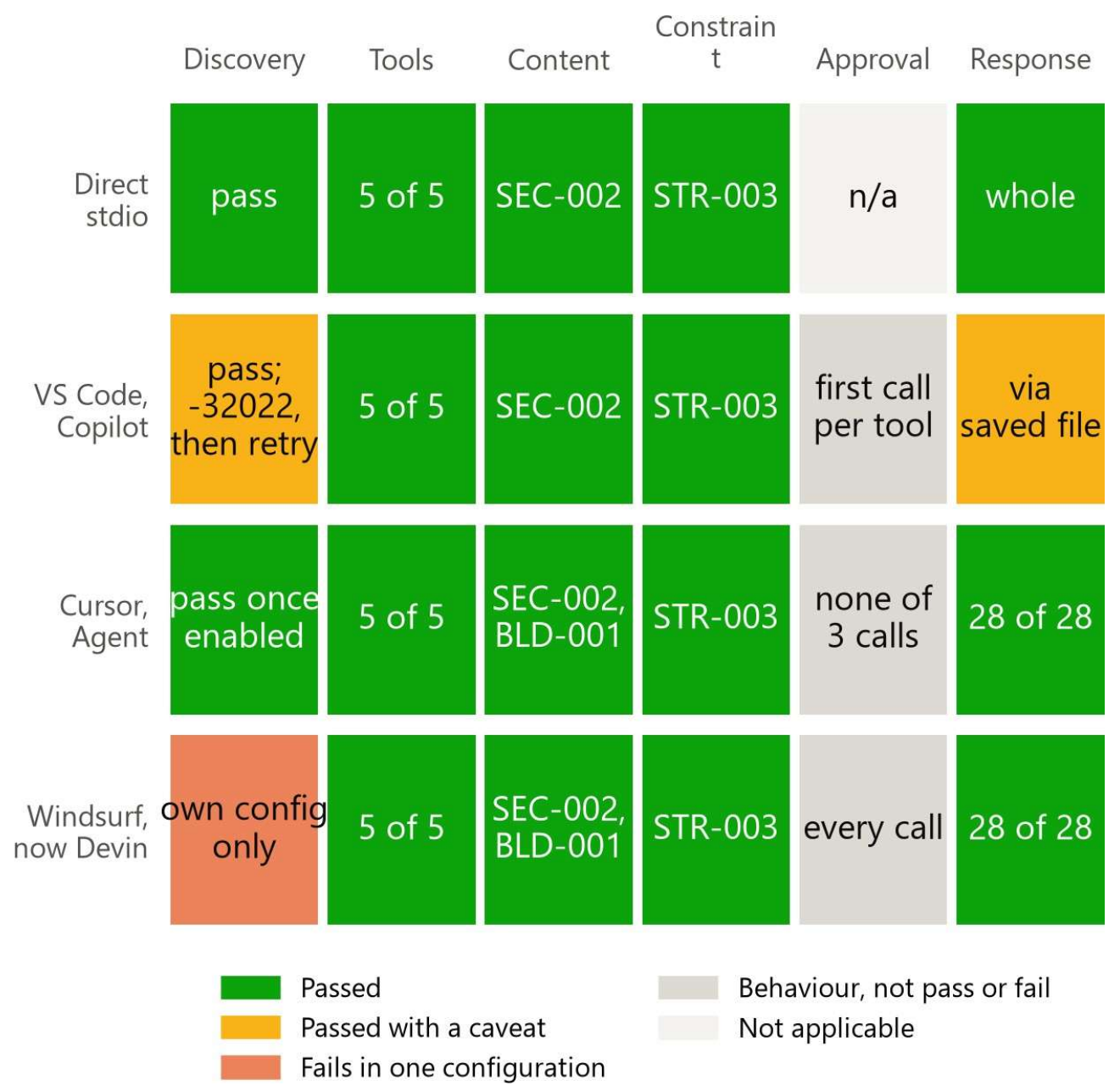

Fig. 16. Compatibility matrix observed on 14 September 2026 for a direct stdio client and for the Copilot agent in VS Code 1.137.0, for Cursor 3.20.17 and for Windsurf 3.10.23, now named Devin. Cell text is abbreviated; the observations are described in the text.

The nine steps completed in 82 seconds (Table XIII). An independent check afterwards found the health path, a versioned API route and the metrics endpoint answering with status 200, every response carrying HSTS, a content security policy, a referrer policy, CORS and rate-limit headers, and frame and MIME-type protection, and no secret in the repository. The workflow then redeployed the service for two later commits and passed its health check with no manual step.

TABLE XIII
END-TO-END DEPLOYMENT

| Step | Result |
|---|---|
| Scoring gate | Score 100, no P0 failure, estimate 0.842 |
| Pre-flight | 4 of 4 checks passed |
| Environment sealing | 2 values sealed |
| Docker build test | Image built, container healthy |
| Git push | 6 generated files pushed |
| Render deployment | Web service created |
| Deploy monitoring | Live after 4 polls |
| Smoke test | 5 of 5 checks passed |
| CI/CD wiring | 3 secrets set, workflow active |

## I. Cross-IDE Compatibility

The server was exercised from a direct stdio client and from the agents of three IDEs, each configured by its own documented method, and every result was read from the client's own logs (Fig. 16). In all of them, once the server started, the five tools were listed and both contract forms, a static content contract and a delegated constraint contract, returned intact; no file of the samples changed. The differences lie around the protocol rather than in it. VS Code's agent refused its first connection with a protocol-version error and connected on retry, a documented negotiation behaviour of the MCP library that the library's own client recovers from. The largest response, the 11,988 byte Express blueprint, exceeded the VS Code agent's limit for inline results and was delivered to the model by way of a file the client saved, while Cursor and Devin passed it inline. Cursor asked for no approval, VS Code asked once for each tool and allowed that approval to cover the session, and Devin asked on every call. Approval is a client setting rather than a fixed property: in a later session, on 16 September 2026 and with a different model selected, Cursor asked once before its first call and not again. Devin blanked the workspace variable in the committed configuration files and connected only through its own local file. No IDE-specific code was added to the server to reach these results.

## J. Research Questions Revisited

**RQ1.** Classification removes the model from 44 of the 50 rules, and the deterministic classes stay accurate: 87.6% of applied static contracts and 95.0% of applied parametric contracts passed their checker, rising to 89.8% for static contracts after contracts began declaring their environment keys, and every shortfall has a stated cause. The delegated class was measured separately with live agents, which resolved all six delegated rules on the first attempt over ten trials, inside the contract's boundary and without regressions.

**RQ2.** Across 499 real repositories the audit finds 7407 violations, and 93.5% of them carry a static or parametric fix, so 6.5% would need model-authored content. Over the sample corpus the figure was 96.7% of the rule instances the loop attempted.

**RQ3.** Independent verification rejected 25 of 237 fixes reported as applied (10.5%) that trusting the report would have accepted, four of which had broken another rule. The reporter was a stand-in executor rather than a language model.

**RQ4.** A model trained on real deployment outcomes ranks deployability at 0.909 ROC AUC, well above the baselines, and rule compliance was shown to differ from deployment outcome. A model trained on compliance labels was not built for direct comparison.

**RQ5.** The unchanged server worked from VS Code, Cursor and Windsurf, with both contract forms intact in each. Launch configuration, approval behaviour and response size differ between clients and must be handled in each client's configuration rather than in the server.

### *K. Threats to Validity*

The reliability figures rest on planted violations in sample projects rather than on a sample of real defects, and several samples were built to be ambiguous, although the split of violations by fix type is measured over real repositories. Delegated fixes by live agents were observed once per rule on one day with two models, which is an existence result and not a rate, the credential rotation that the hygiene rules depend on happens outside the repository and was not performed, no agent has yet invoked the deployment tool, and in the full-chain run the external services behind stages 5 to 8 were scripted; the single live deployment covers the Express path. The build pass behind the model's key feature was run about a week later than the deployments, and three or more projects built differently from Render; 39 rows were built on Node 24 for want of an older image, and without them the held-out ROC AUC is 0.898. For React, a local build is nearly the same operation Render performs to decide the label, which partly explains how well React separates. The study covers two stacks and one provider, and the ruleset was written by the same work that evaluates against it.

## VIII. Conclusion

Automated production readiness repair is usually framed as generation followed by checking. ProdPilot shows that most of the domain needs no generation: 44 of 50 rules are repaired from templates or extracted values, and the six that need judgement are written by the IDE agent within a boundary. Trust is replaced by verification, since every fix is accepted only if the flagging checker passes, and a fix that breaks another rule is reverted; across 23 sample projects this rejected 25 of 237 fixes that had been reported as applied. Across 499 real repositories, 93.5% of the violations the audit finds carry a fix that needs no model. The release decision joins the audit with a deployability model learned from 684 real deployments, a repaired project reached a live, smoke-tested service with a working pipeline in 82 seconds, and the unchanged server worked from the agents of three IDEs. The remaining work is measurement: delegated fixes at a scale that yields a rate rather than an existence result, and a compliance-labelled baseline for the model.